\documentclass[pdftex,twocolumn,epjc3,nofootinbib,amsmath,amssymb]{svjour3}          % twocolumn

\RequirePackage[T1]{fontenc}

\RequirePackage{graphicx}
\RequirePackage{mathptmx}      % use Times fonts if available on your TeX system
\RequirePackage{flushend}
\RequirePackage[numbers,sort&compress]{natbib}
\RequirePackage[colorlinks,citecolor=blue,urlcolor=blue,linkcolor=blue, draft]{hyperref}
\usepackage{bm}
\usepackage{amsfonts,amssymb,amsmath,mathtools,physics}% math and phys
\usepackage[subnum]{cases}% adds the numcases environment
\usepackage{enumitem}% add roman letters enumerate
\usepackage{soul}

\journalname{Eur. Phys. J. C}
\newcommand{\Sh}{Schr\"{o}dinger}
\newcommand{\numberset}{\mathbb}
\newcommand{\R}{\numberset{R}}
\newcommand{\N}{\numberset{N}}
\newcommand{\Z}{\numberset{Z}}
\newcommand{\W}{\mathcal{W}}
\newcommand{\Hi}{\mathcal{H}}

\newcommand{\C}{\mathcal{C}}
\newcommand{\half}{\frac{1}{2}}

\begin{document}

\title{Semiclassical and quantum behavior of the Mixmaster model in  the polymer 
        approach for the isotropic Misner variable}


\author{C. Crin\`o\thanksref{e1,addr1}
        \and
        G. Montani\thanksref{e2,addr1,addr2}
        \and
        G. Pintaudi \thanksref{e3,addr1} %etc.
}

\thankstext{e1}{e-mail: crino.1547608@studenti.uniroma1.it}
\thankstext{e2}{e-mail: giovanni.montani@frascati.enea.it}
\thankstext{e3}{e-mail: pintaudi.1274530@studenti.uniroma1.it}

\institute{Dipartimento di Fisica (VEF), P.le A. Moro 5 (00185)
  Roma, Italy\label{addr1}
          \and
         ENEA, Fusion and Nuclear Safety Department, C.R. Frascati, via E. Fermi 45, 00044
  Frascati (Roma), Italy\label{addr2}
}

\date{\today}%Received: date / Accepted: date

\maketitle

\begin{abstract}
 We analyze the semi-classical and quantum behavior of the Bianchi IX
  Universe in the Polymer Quantum Mechanics framework, applied to the
  isotropic Misner variable, linked to the space volume of the model.
  The study is performed both in the Hamiltonian and field equations
  approaches, leading to the remarkable result of a still singular and
  chaotic cosmology, whose Poincar\'e return map asymptotically
  overlaps the standard Belinskii-Khalatnikov-Lifshitz one.  In the
  quantum sector, we reproduce the original analysis due to Misner,
  within the revised Polymer approach and we arrive to demonstrate
  that the quantum numbers of the point-Universe still remain
  constants of motion. This issue confirms the possibility to have
  quasi-classical states up to the initial singularity.  The present
  study clearly demonstrates that the asymptotic behavior of the
  Bianchi IX Universe towards the singularity is not significantly
  affected by the Polymer reformulation of the spatial volume dynamics
  both on a pure quantum and a semiclassical level.
\end{abstract}

\section{Introduction}

The Bianchi IX Universe~\cite{Landau:1980,Misner:1973} is the most
interesting among the Bianchi models. In fact, like the Bianchi type
VIII, it is the most general allowed by the homogeneity constraint,
but unlike the former, it admits an isotropic limit, naturally reached
during its evolution \cite{Grishchuk:1975, Doroshkevich:1973} (see
also \cite{Kirillov:2002,Montani:2008}), coinciding with the
positively curved Robertson-Walker geometry. Furthermore, the
evolution of the Bianchi IX model towards the initial singularity is
characterized by a chaotic structure, first outlined
in~\cite{Belinskii:1970} in terms of the Einstein equations morphology
and then re-analyzed in Hamiltonian formalism by \cite{Misner:1969,
  Misner:1969mixmaster}.  Actually, the Bianchi IX chaotic evolution
towards the singularity constitutes, along with the same behavior
recovered in Bianchi type VIII, the prototype for a more general
feature, characterizing a generic inhomogeneous cosmological model
\cite{Belinskii:1982} (see also \cite{Kirillov:1993, Montani:1995,
  Montani:2008, Montani:2011,Heinzle:2009, Barrow:1979}).

The original interest toward the Bianchi IX chaotic cosmology was due
to the perspective, de facto failed, to solve the horizon paradox via
such a statistical evolution of the Universe scale factors, from which
derives the name, due to C. W. Misner, of the so-called
\emph{Mixmaster model}. However, it was clear since from the very
beginning, that the chaotic properties of the Mixmaster model had to
be replaced, near enough to the initial singularity (essentially
during that evolutionary stage dubbed Planck era of the Universe), by
a quantum evolution of the primordial Universe.  This issue was first
addressed in~\cite{Misner:1969}, where the main features of the
Wheeler-De Witt equation~\cite{DeWitt:1967} are implemented in the
scenario of the Bianchi IX cosmology. This analysis, besides its
pioneering character (see also~\cite{Graham:1994, Montani:2008}),
outlined the interesting feature that a quasi-classical state of the
Universe, in the sense of large occupation numbers of the
wave function, is allowed up to the initial singularity.

More recent approaches to Canonical Quantum Gravity, like the
so-called Loop Quantum Gravity~\cite{Cianfrani:2014}, suggested that
the geometrical operators, areas and volumes, are actually
characterized by a discrete spectrum~\cite{Rovelli:1995}.  The
cosmological implementation of this approach led to the definition of
the concept of ``Big-Bounce''~\cite{Ashtekar:2006quantum,
  Ashtekar:2006num} (i.e.\ to a geometrical cut-off of the initial
singularity, mainly due to the existence of a minimal value for the
Universe Volume) and could transform the Mixmaster Model in a cyclic
Universe~\cite{Montani:2018,Barrow:2017}.  The complete implementation
of the Loop Quantum Gravity to the Mixmaster model is not yet
available~\cite{Ashtekar:2009, Ashtekar:2011bkl, Bojowald:2004ra}, but
general considerations on the volume cut-off led to characterize the
semi-classical dynamics as chaos-free~\cite{Bojowald:2004}. A
quantization procedure, able to mimic the cut-off physics contained in
the Loop Quantum Cosmology, is the so-called \emph{Polymer quantum
  Mechanics}, de facto a quantization on a discrete lattice of the
generalized coordinates~\cite{Ashtekar:2001xp} (for cosmological
implementations see~\cite{Corichi:2007pr, Montani:2013, Moriconi:2017,
  Battisti:2008, Hossain:2010, DeRisi:2007, Hassan:2017,
  Seahra:2012}).

Here, we apply the Polymer Quantum Mechanics to the isotropic variable
$\alpha$ of the Mixmaster Universe, described both in the Hamiltonian
and field equations representation.  We first analyze the Hamiltonian
dynamics in terms of the so called semiclassical polymer equations,
obtained in the limit of a finite lattice scale, but when the
classical limit of the dynamics for $\hbar\rightarrow 0$ is taken.
Such a semiclassical approach clearly offers a characterization for
the behavior of the mean values of the quantum Universe, in the spirit
of the Ehrenfest theorem~\cite{Sakurai:2011}.  This study 
demonstrates that the singularity is not removed and the chaoticity of
the Mixmaster model is essentially preserved, and even enforced in
some sense, in this cut-off approach.

Then, in order to better characterize the chaotic features of the
obtained dynamics, we translate the Hamiltonian polymer-like dynamics,
in terms of the modified Einstein equations, in order to calculate the
morphology of the new Poincar\'e return map, i.e.\ the modified BKL
map.

We stress that, both the reflection rule of the point-Universe against
the potential walls and the modified BKL map acquire a readable form
up to first order in the small lattice step parameter. Both these two
analyses clarify that the chaotic properties of the Bianchi IX model
survive in the Polymer formulation and are expected to preserve the
same structure of the standard General Relativity case.  In
particular, when investigating the field equations, we numerically
iterate the new map, showing how it asymptotically overlaps to the
standard BKL one.

The main merit of the present study is to offer a detailed and
accurate characterization of the Mixmaster dynamics when the Universe
volume (i.e.\ the isotropic Misner variable) is treated in a
semiclassical Polymer approach, demonstrating how this scenario does
not alter, in the chosen representation, the existence of the initial singularity and the chaoticity
of the model in the asymptotic neighborhoods of such a singular point.

Finally, we repeat the quantum Misner analysis in the Polymer revised
quantization framework and we show, coherently with the semiclassical
treatment, that the Misner conclusion about the states with high
occupation numbers, still survives: such occupation numbers still
behave as constants of motion in the asymptotic dynamics.

It is worth to be noted that imposing a discrete structure to the
variable $\alpha$ does not ensure that the Bianchi IX Universe volume
has a natural cut-off. In fact, such a volume behaves like
$e^{3\alpha}$ and it does not take a minimal (non-zero) value when
$\alpha$ is discretized.  This fact could suggest that the surviving
singularity is a consequence of the absence of a real geometrical
cut-off, like in Loop Quantum Cosmology (of which our approach mimics
the semi-classical features).  However, the situation is subtler, as
clarified by the following two considerations.
\begin{enumerate}[label= (\roman*)]
\item In~\cite{Montani:2013}, where the polymer approach is applied to
  the anisotropy variables $\beta _{\pm}$, the Mixmaster chaos is removed 
 like in \cite{Bojowald:2004} even if no discretization is imposed on the isotropic (volume)
  variable.  Furthermore, approaching the singularity, the
  anisotropies classically diverge and their discretization does not
  imply any intuitive regularization, as de facto it takes place
  there.  The influence of the discretization induced by the Polymer
  procedure, can affect the dynamics in a subtle manner,
  non-necessarily predicted by the simple intuition of a cut-off
  physics, but depending on the details of the induced symplectic
  structure.
\item In Loop Quantum Gravity, the spectrum of the geometrical spatial
  volume is discrete, but it must be emphasized that such a spectrum
  still contains the zero eigenvalue.  This observation, when the
  classical limit is constructed including the suitably weighted zero
  value, allows to reproduce the continuum properties of the classical
  quantities.  Such a point of view is well-illustrated
  in~\cite{Rovelli:2002vp}, where the preservation of a classical
  Lorentz transformation is discussed in the framework of Loop Quantum
  Gravity.  This same consideration must hold in Loop Quantum
  Cosmology, for instance in~\cite{Ashtekar:2006quantum}, where the
  position of the bounce is determined by the scalar field parameter,
  i.e., on a classical level, it depends also on the initial
  conditions and not only on the quantum modification to geometrical
  properties of space-time.
\end{enumerate}
Following the theoretical paradigm suggested by the point (ii), the
analysis of the Polymer dynamics of the Misner isotropic variable
$\alpha$ is extremely interesting because it corresponds to a
discretization of the Universe volume, but allowing for the value zero
of such a geometrical quantity.

We note that picking, as the phase-space variable to quantize, the
scale factor $a = e^{\alpha}$, or any other given power of this, the
corresponding Polymer discretization is somehow forced to become
singularity-free, i.e.\ we observe the onset of a Big Bounce. However,
in the present approach, based on a logarithmic scale factor, the
polymer dynamical scheme offers an intriguing arena to test
the Polymer Quantum Mechanics in the Minisuperspace, even in
comparison with the predictions of Loop Quantum
Cosmology~\cite{Cianfrani:2014}.

Finally, we are interested in determining the fate of the Mixmaster
model chaoticity, which is suitably characterized by the Misner
variables representation and whose precise description (i.e.\
ergodicity of the dynamics, form of the invariant measure) is reached
in terms of the Misner-Chitr\'e-like variables~\cite{Montani:2011,
  Montani:1997, Imponente:2001fy}, which are double logarithmic scale
factors. Although these variables mix the isotropic and anisotropic
Misner variables to some extent, the Misner-Chitr\'e time-like variable in the
configurational scale, once discretized would not guarantee a minimal value
of the universe volume.
Thus the motivations for a Polymer treatment
of the Mixmaster model in which only the $\alpha$ dynamics is
deformed relies on different and converging requests, coming from the
cosmological, the statistical and the quantum feature of the Bianchi
IX universe.

The paper is organized as follows. In Section \ref{sec:PQM} we
introduce the main kinematic and dynamic features of the Polymer
Quantum Mechanics. Then we outline how this singular representation
can be connected with the Schr\"odinger one through an appropriate
Continuum Limit.

In Section \ref{sec:mixm-model:-class} we describe the dynamics of the
homogeneous Mixmaster model as it can be derived through the Einstein
field equations. A particular attention is devoted to the Bianchi I
and II models, whose analysis is useful for the understanding of the
general Bianchi IX (Mixmaster) dynamics.

The Hamiltonian formalism is introduced in Section
\ref{sec:HamiltonianFormalism}, where we review the semiclassical and
quantum properties of the model as it was studied by Misner in
\cite{Misner:1969}.

Section \ref{sec:mixm-poly} is devoted to the analysis of the polymer
modified Mixmaster model in the Hamiltonian formalism, both from a
semiclassical and a quantum point of view. Our results are then
compared with the ones derived by some previous models.

In Section~\ref{sec:polymer-bkl-map} the semiclassical
behavior of the polymer modified Bianchi I and Bianchi II models is
developed through the Einstein equations formalism, while the modified
BKL map is derived and its properties discussed from an analytical and
numerical point of view in Section~\ref{sec:polymer-BKL-map}.

In Section \ref{sec:physical-considerations} we discuss two important physical issues, concerning
the link of the polymer representation with Loop Quantum Cosmology and the implications
of a polymer quantization of the whole Minisuperspace variables on the Mixmaster chaotic features.

Finally, in Section \ref{sec:conclusions} brief concluding remarks follow.

\section{Polymer Quantum Mechanics}\label{sec:PQM}

The Polymer Quantum Mechanics (PQM) is a representation of the usual
canonical commutation relations (CCR), unitarily nonequivalent to the
\Sh\ one.  It is a very useful tool to investigate the consequences of
the assumption that one or more variables of the phase space are
discretized. Physically, it accounts for the introduction of a
cutoff. In certain cases where the \Sh\ representation is
well-defined, the cutoff can be removed through a certain limiting
procedure and the usual \Sh\ representation is recovered, as shown in
\cite{Corichi:2007pr} and summed up in Sec.~\ref{sec:continuumLimit}.

Many people in the Quantum Gravity community think that there is a
maximum theoretical precision achievable when measuring space-time
distances, this belief being backed up by valuable although heuristic
arguments
\cite{Salecker-Wigner:1958,AmelinoCamelia:1999-Salecker-Wigner,
  Hossenfelder:2012jw}, so that the cutoff introduced in PQM is
assumed to be a fundamental quantity.  Some results of Loop Quantum
Gravity\cite{Ashtekar:2011} (the discrete spectrum of the area
operator) and String Theory\cite{Lidsey:1999} (the minimum length of a
string) point clearly in the direction of a minimal length scale
scenario, too.

PQM was first developed by Ashtekar et
al.\cite{Ashtekar:2001xp,Ashtekar:2002,Ashtekar:2003} who also credit
a previous work of Varadarajan\cite{Varadarajan:2000} for some
ideas. It was then further refined also by Corichi et
al.\cite{Corichi:2007pr,Corichi:2007cqg}. They developed the PQM in
the expectation to shed some light on the connection between the
Planckian-energy Physics of Loop Quantum Gravity and the lower-energy
Physics of the Standard Model.

\subsection{The \Sh\ representation}

Let us consider a quantum one-dimensional free particle with phase
space $(q,p) \in \Gamma = \R^2$. The standard CCR are summarized by
the relation
\begin{equation}\label{standard-CCR}
  \left[ \hat{q}, \hat{p} \right] = i \hat{I}
\end{equation}
where $\hat{q}$ and $\hat{p}$ are the position and momentum operators
respectively and $\hat{I}$ is the identity operator. These operators
form a basis for the so called Heisenberg algebra\cite{Binz:2008}. At
this stage we have not made any assumption regarding the Hilbert space
of the theory, yet.

To introduce the Polymer representation in Sec.~\ref{sec:kinematics},
it is convenient to consider also the Weyl algebra $\W$, generated by
exponentiation of $\hat{q}$ and $\hat{p}$
\begin{equation}\label{UV-basis}
  \hat{U}(\alpha_i) = e^{i\alpha\hat{q}}; \quad \hat{V}(\beta_i) = 
  e^{i\beta\hat{p}}
\end{equation}
where $\alpha$ and $\beta$ are real parameters with the dimension of
momentum and length, respectively.

The CCR~\eqref{standard-CCR} become
\begin{equation}\label{Weyl-CCR}
  \hat{U}(\alpha) \cdot \hat{V}(\beta) =
  e^{-i\alpha\beta}\hat{V}(\beta) \cdot \hat{U}(\alpha)
\end{equation}
where $\cdot$ indicates the product operation of the algebra.  All
other product combinations commute.

A generic element of the Weyl algebra $W \in \W$ is a finite linear
combination
\begin{equation}
  W(\alpha, \beta) = \sum_i (A_i U(\alpha_i) + B_i V(\beta_i))
\end{equation}
where $A_i$ and $B_i$ are complex coefficients and the
product~(\ref{Weyl-CCR}) is extended by linearity. It can be shown
that $\W$ has a structure of a $\C^{*}$-algebra.

The \Sh\ representation is obtained as soon as we introduce the \Sh\
Hilbert space of square Lebesgue-integrable functions
$\Hi_S = L^2(\R,dq)$ and the action of the bases operators on it:
\begin{equation}
  \hat{q} \psi(q) = q \psi(q); \quad \hat{p} \psi(q) = -i\partial_q
  \psi(q)
\end{equation}
where $\psi \in L^2(\R,dq)$ and $q \in \R$.

\subsection{The polymer representation:
  kinematics}\label{sec:kinematics}

The Polymer representation of Quantum Mechanics can be introduced as
follows. First, we consider the abstract states $\ket{\mu}$ of the
Polymer Hilbert space $\Hi_{\text{poly}}$ labeled by the real
parameter $\mu$. A generic state of $\Hi_{\text{poly}}$ consists of a
finite linear combination
\begin{equation}\label{polymer-generic-ket}
  \ket{\Psi} = \sum^N_{i = 1} a_i \ket{\mu_i}
\end{equation}
where $N\in\N$ and we assume the fundamentals kets to be orthonormal
\begin{equation}\label{polymer-inner-product}
  \braket{\mu}{\nu} = \delta_{\mu,\nu}
\end{equation}
The Hilbert space $\Hi_{\text{poly}}$ is the Cauchy completion of the
vectors~\eqref{polymer-generic-ket} with respect to the
product~\eqref{polymer-inner-product}. It can be shown that
$\Hi_{\text{poly}}$ is nonseparable\cite{Corichi:2007pr}.

Two fundamental operators can be defined on this Hilbert space.  The
``label'' operator $\hat{\epsilon}$ and the ``displacement'' operator
$\hat{\boldsymbol{s}}(\lambda)$ with $\lambda \in \R$, which act on
the states $\ket{\mu}$ as
\begin{equation}
  \hat{\epsilon}\ket{\mu} \coloneqq \mu\ket{\mu}; \quad
  \hat{\boldsymbol{s}}(\lambda)\ket{\mu} \coloneqq \ket{\mu + \lambda}
\end{equation}
The $\hat{\boldsymbol{s}}$ operator is discontinuous in $\lambda$
since for each value of $\lambda$ the resulting states
$\ket{\mu + \lambda}$ are orthogonal. Thus, there is no Hermitian
operator that by exponentiation could generate the displacement
operator.

We denote the wave functions in the $p$-polarization as
$\psi(p) = \braket{p}{\psi}$, where
$\psi_\mu(p) = \braket{p}{\mu} = e^{i\mu p}$.  The $\hat{V}(\lambda)$
operator defined in~\eqref{UV-basis} shifts the plane waves by
$\lambda$
\begin{equation}\label{V-as-shift-operator}
  \hat{V}(\lambda) \cdot \psi_\mu(p) = e^{i\lambda p} e^{i\mu p} =
  e^{i(\lambda + \mu) p} = \psi_{(\lambda + \mu)}(p)
\end{equation}
As expected, $\hat{V}(\lambda)$ can be identified with the
displacement operator $\hat{\boldsymbol{s}}(\lambda)$ and the
$\hat{p}$ operator cannot be defined rigorously. On the other hand the
operator $\hat{q}$ is defined by
\begin{equation}
  \hat{q} \cdot \psi_\mu(p) = -i \frac{\partial}{\partial p} \psi_\mu(p)
  = \mu e^{i\mu p} = \mu \psi_\mu(p)
\end{equation}
and thus can be identified with the abstract displacement operator
$\hat{\epsilon}$. The reason $\hat{q}$ is said to be discrete is that
the eigenvalues of this operator are the labels $\mu$, and even when
$\mu$ can take value in a continuum of possible values, they can be
regarded as a discrete set, because the states are orthonormal for all
values of $\mu$.

The $\hat{V}(\lambda)$ operators form a $\C^{*}$-Algebra and the
mathematical theory of $\C^{*}$-Algebras\cite{Arveson:1998} provide us
with the tools to characterize the Hilbert space which the wave
functions $\psi(p)$ belong to. It is given by the \emph{Bohr
  compactification} $\R_b$ of the real line\cite{Rudin:2011}:
$\Hi_{\text{poly}} = L^2(\R_b,d\mu_H)$, where the Haar measure
$d\mu_H$ is the natural choice.

In terms of the fundamental functions $\psi_\mu(p)$ the inner product
takes the form
\begin{equation}\label{abstract-PQM-inner-product}
  \begin{split}
    \braket{\psi_\mu}{\psi_\nu} \coloneqq & \int_{\R_b} d\mu_H
    \bar{\psi_\mu}(p) \psi_\nu(p) \\ \coloneqq &\lim_{L\to\infty}
    \frac{1}{2L} \int^L_{-L} dp \bar{\psi_\mu}(p) \psi_\nu(p) =
    \delta_{\mu,\nu}
  \end{split}
\end{equation}

\subsection{Dynamics}\label{sec:dynamics}

Once a definition of the polymer Hilbert space is given, we need to
know how to use it in order to analyze the dynamics of a system. The
first problem to be faced is that, as it was said after equation
\eqref{V-as-shift-operator}, the polymer representation isn't able to
describe $\hat{p}$ and $\hat{q}$ at the same time. It is then
necessary to choose the ``discrete'' variable and to approximate its
conjugate momentum with a well defined and quantizable function.  For
the sake of simplicity let us investigate the simple case of a
one-dimensional particle described by the Hamiltonian:
\begin{equation}
  H=\frac{p^2}{2m}+\mathcal{V}(q)
  \label{chiara:ParticleHam}
\end{equation}
in p-polarization.  If we assume that $\hat{q}$ is the discrete
operator (in the sense explained in Sec. \ref{sec:kinematics}), we
then need to approximate the kinetic term $\frac{p^2}{2m}$ in an
appropriate manner. The procedure outlined in \cite{Corichi:2007pr}
consists in the regularization of the theory, through the introduction
of a regular graph $\gamma_{\mu_0}$ (that is a lattice with spacing
$\mu_0$), such as:

\begin{equation}
  \gamma_{\mu_0}=\{q\in\mathbb{R}\ |\ q=n\mu_0,\ \forall n\in\mathbb{Z}\}
\end{equation} 
It follows that the only states we shall consider, in order to remain
in the graph, are that of the form
$\ket{\psi}=\sum_n{b_n\ket{\mu_n}}\in \mathcal{H}_{\gamma_{\mu_0}}$,
where the $b_n$ are coefficients such as $\sum_n|b_n|^2<\infty$ and
the $\ket{\mu_n}$ (with $\mu_n=n\mu_0$) are the basic kets of the
Hilbert space $\mathcal{H}_{\gamma_{\mu_0}}$, which is a separable
subspace of $\mathcal{H}_{poly}$.  Moreover the action of all
operators on these states will be restricted to the graph.  Since the
exponential operator $\hat{V}(\lambda)=e^{i\lambda p}$ can be
identified with the displacement one $\hat{s}$ (Eq.
\eqref{V-as-shift-operator}), it is possible to use a regularized
version of it
($\hat{V}(\mu_0)$ such that
$\hat{V}(\mu_0)\ket{\mu_n}=\ket{\mu_n+\mu_0}=\ket{\mu_{n+1}}$) in order
to define an approximated version of $\hat{p}$:
\begin{multline}
  \hat{p}_{\mu_0}\ket{\mu_n}=\frac{1}{2i\mu_0}[\hat{V}(\mu_0)-\hat{V}(-\mu_0)]\ket
  {\mu_n}=\\
  =-\frac{i}{2\mu_0}(\ket{\mu_{n+1}}-\ket{\mu_{n-1}})
\end{multline}
This definition is based on the fact that, for $p<<\frac{1}{\mu_0}$,
one gets
$p\simeq\frac{1}{\mu_0}\sin(\mu_0 p)=\frac{1}{2i\mu_0}(e^{i\mu_0
  p}-e^{-i\mu_0 p})$.  From this result it is easy to derive the
approximated kinetic term $\hat{p}_{\mu_0}^2$ by applying the same
definition of $\hat{p}_{\mu_0}$ two times:
\begin{multline}
  \hat{p}_{\mu_0}^2\ket{\mu_n}=\hat{p}_{\mu_0}\cdot\hat{p}_{\mu_0}\ket{\mu_n}=\\
  =\frac{1}{4\mu_0^2}(\ket{\mu_{n+2}}+\ket{\mu_{n-2}}-2\ket{\mu_n})
\end{multline}
It is now possible to introduce a regularized Hamiltonian operator,
which turns out to be well-defined on $\mathcal{H}_{\gamma_{\mu_0}}$
and symmetric:
\begin{equation}
  \hat{H}_{\mu_0}=\frac{\hat{p}_{\mu_0}^2}{2m}+\mathcal{V}(\hat{q})\in\mathcal{H}_
  {\gamma_0}
\end{equation}
In the p-polarization, then, $\hat{p}^2_{\mu_0}$ acts as a
multiplication operator
($\hat{p}^2_{\mu_0}\psi(p)=\frac{1}{\mu_0^2}\sin^2(\mu_0 p)\psi(p)$),
while $\hat{q}$ acts as a derivative operator
($\hat{q}\psi(p)=i\partial_p\psi(p)$).

\subsection{Continuum Limit}\label{sec:continuumLimit}

As we have seen in the previous section, the condition
$p<<\frac{1}{\mu_0}$ is a fundamental hypothesis of the polymer
approach. If this limit is valid for all the orbits of the system to
be quantized, such a representation is expected to give results
comparable to the ones of the standard quantum representation (based
on the Schr\"odinger equation). This is the reason why a limit
procedure from the polymer system to the Schr\"odinger one is
necessary to understand how the two approaches are related. In other
words we need to know if, given a polymer wave function defined on the
graph $\gamma_0$ (with spacing $a_0$), it is possible to find a
continuous function which is best approximated by the previous one in
the limit when the graph becomes finer, that is:
\begin{equation}
  \gamma_0\to\gamma_n=\left\{q_k\in\mathbb{R}\ |\ q_k=ka_n,\ \text{with}\ 
    a_n=\frac{a_0}{2^n},\ \forall k\in\mathbb{Z}\right\}
\end{equation}
The first step to take in order to answer this question consists in
the introduction of the scale $C_n$ which is a decomposition of the
real line in the union of closed-open intervals with the lattice
$\gamma_n$ points as extrema, that cover the whole line and do not
intersect. For each scale we can then define an effective theory based
on the fact that every continuous function can be approximated by
another function that is constant on the intervals defined by the
lattice. It follows that we can create a link between
$\mathcal{H}_{poly}$ and $\mathcal{H}_S$ for each scale $C_n$:
\begin{equation}
  \sum_m\psi(ma_n)\delta_{ma_n,q}\in\mathcal{H}_{poly}\to\sum_m\psi(ma_n)\chi_{
    \alpha_m}(q)\in\mathcal{H}_S
\end{equation}
where $\chi_{\alpha_m}$ is the characteristic function on the interval
$\alpha_m=[ma_n,(m+1)a_n)$.  Thus it is possible to define an
effective Hamiltonian $\hat{H}_{C_n}$ for each scale $C_n$. The set of
effective theories is then analyzed by the use of a renormalization
procedure \cite{Corichi:2007pr}, and the result is that the existence
of a continuum limit is equivalent to both the convergence of the
energy spectrum of the effective theory to the continuous one and the
existence of a complete set of renormalized eigenfunctions.

\section{The homogeneous Mixmaster model: classical dynamics}
\label{sec:mixm-model:-class}

With the aim of investigating in Sec.~\ref{sec:polymer-BKL-map} the
modifications to the Bianchi IX dynamics produced by the introduction
of the Polymer cutoff, in this Section we briefly review some relevant
results obtained by Belinskii, Khalatnikov and Lifshitz (BKL)
regarding the classical Bianchi IX model.  In particular they found
that an approximate solution for the Bianchi IX dynamics can be given
in the form of a Poincar\'e recursive map called \emph{BKL map}. This
map is obtained as soon as the exact solutions of Bianchi I and
Bianchi II are known. Hence, we will first derive the dynamical
solution to the classical Bianchi I and II models.

The Einstein equations of a generic Bianchi model can be expressed in
terms of the \emph{spatial metric}
$\eta(t)=\text{diag}[a(t),\\ b(t),c(t)]$ and the \emph{constants of
        structure} $\lambda_l,\lambda_m,\lambda_n$ of the inequivalent
isometry group that characterize each Bianchi model, where ``$a,b,c$''
are said \emph{cosmic scale factors} and describe the behavior in the
synchronous time $t$ of the three independent spatial
directions\cite{Landau:1980}.
\begin{subequations}\label{gener-dynam-bianchi:log-variables}
        \begin{align}
        (q_l)_{\tau\tau} = & \frac{1}{2a^2 b^2 c^2}
        \left ( {\lambda_l}^2 a^4 - {\left (
                \lambda_m b^2 -
                \lambda_n c^2 \right )}^2 \right )\\
        (q_m)_{\tau\tau} = & \frac{1}{2a^2 b^2 c^2}
        \left ( {\lambda_m}^2 b^4 - {\left (
                \lambda_l a^2 -
                \lambda_n c^2 \right )}^2 \right )\\
        (q_n)_{\tau\tau} = & \frac{1}{2a^2 b^2 c^2}
        \left ( {\lambda_n}^2 c^4 - {\left (
                \lambda_l a^2 -
                \lambda_m b^2 \right )}^2 \right )
        \end{align}
\end{subequations}
\begin{equation}\label{gener-dynam-bianchi:constraint}
{(q_l + q_m + q_n)}_{\tau\tau} = {(q_l)}_\tau {(q_m)}_\tau
+ {(q_l)}_\tau {(q_n)}_\tau + {(q_m)}_\tau {(q_n)}_\tau
\end{equation}
where the logarithmic variable $q_l(\tau),q_m(\tau),q_n(\tau)$ and
logarithmic time $\tau$ are defined as such:
\begin{equation}
q_l = 2\ln a, \quad q_m = 2\ln b, \quad q_n = 2\ln c, \quad dt
= (abc) d\tau
\label{log-variables-definition}
\end{equation}
and the subscript $\Box_{\tau}$ is a shorthand for the derivative in
$\tau$. The isometry group characteristic of Bianchi IX is the SO(3)
group. The Class A Bianchi models\cite{Montani:2011} (of which Bianchi
I,II and IX are members) are set apart just by those three constants of
structure.

\subsection{Bianchi I}\label{sec:bianchi-i}

The Bianchi I model in void is also called \emph{Kasner
        solution}\cite{Wainwright:2008}. By substituting in
Eqs~\eqref{gener-dynam-bianchi:log-variables} the Bianchi I constants
of structure $\lambda_l=0,\lambda_m=0,\lambda_n=0$, we get the simple
equations of motion:
\begin{equation}\label{bianchi-i:equations-of-motion}
{(q_l)}_{\tau\tau} = {(q_m)}_{\tau\tau} = {(q_n)}_{\tau\tau} = 0
\end{equation}
whose solution can be given as a function of the logarithmic time
$\tau$ and four parameters:
\begin{equation}\label{bianchi-i:solution}
\begin{dcases}
q_l(\tau) = 2 \Lambda p_l \tau\\
q_m(\tau) = 2 \Lambda p_m \tau\\
q_n(\tau) = 2 \Lambda p_n \tau
\end{dcases}
\end{equation}
where $p_l,p_m,p_n$ are said \emph{Kasner indices} and $\Lambda$ is a
positive constant that will have much relevance in
Section~\ref{sec:polymer-bianchi-i-dynamics}. It is needed to ensure
that the sum of Kasner indices is always one:
\begin{equation}\label{kasner:sum}
p_l + p_m + p_n = 1
\end{equation}

From~\eqref{bianchi-i:solution}
and~\eqref{log-variables-definition}, if we
exploit the constraint~\eqref{kasner:sum}, we obtain
\begin{equation}\label{bianchi-i:t-tau}
\tau = \frac{1}{\Lambda} \ln(\Lambda t)
\end{equation}
From relation~\eqref{bianchi-i:t-tau} stems the appellative
``logarithmic time'' for the time variable $\tau$.

By substituting the Bianchi I equations of
motion~\eqref{bianchi-i:equations-of-motion} and their
solution~\eqref{bianchi-i:solution}
into~\eqref{gener-dynam-bianchi:constraint}, after applying the
constraint~\eqref{kasner:sum}, we find the additional constraint
\begin{equation}\label{kasner:sumofsquares} 
{p_l}^2 + {p_m}^2 + {p_n}^2 = 1
\end{equation}

\begin{figure}
        \begin{center}
                \includegraphics[width=0.6\columnwidth]{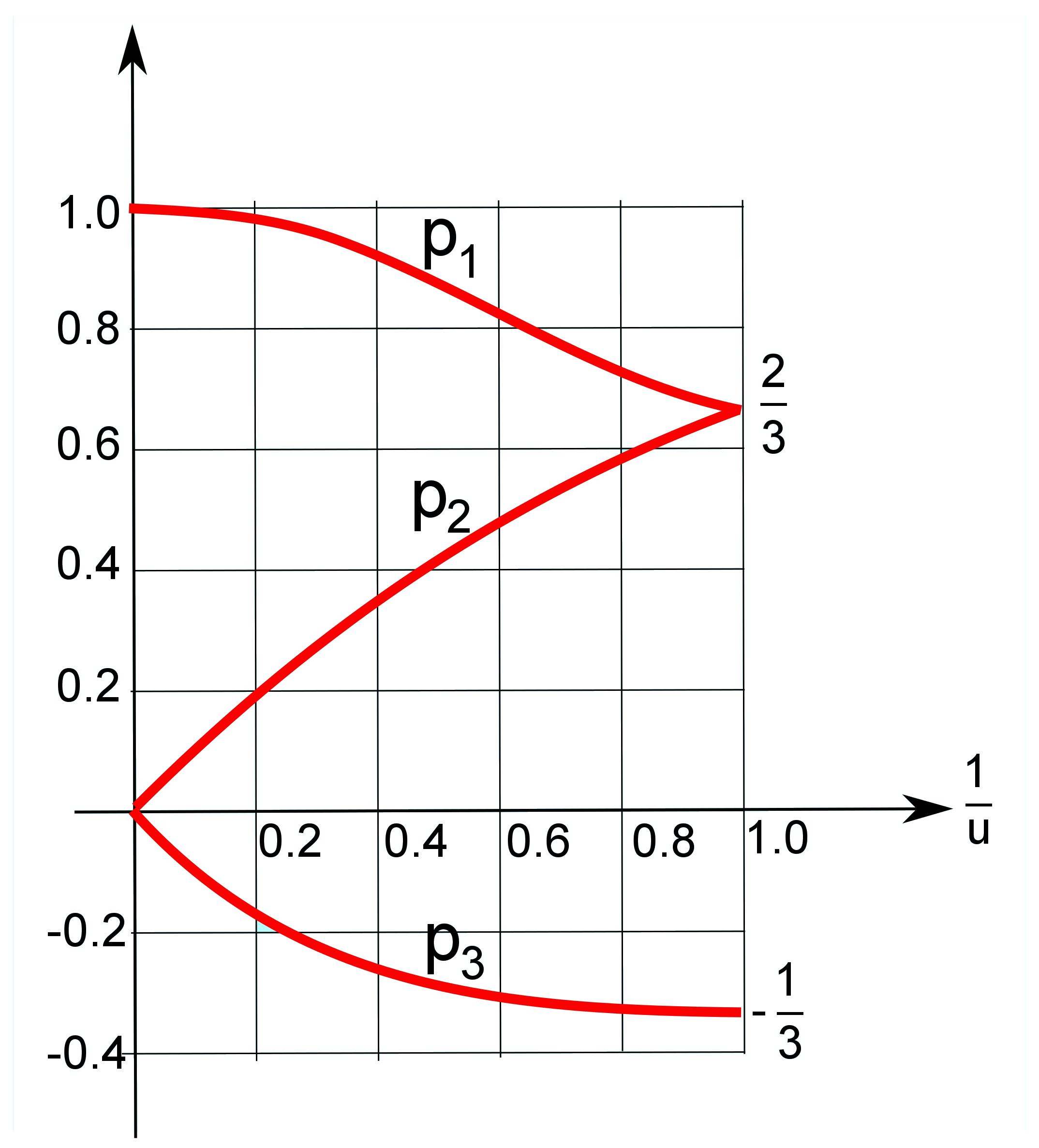}
        \end{center}
        \caption{Kasner indices as functions of the inverse of the $u$
                parameter. This figure was taken from~\cite{wiki:BKL} and is
                released under
                \href{https://creativecommons.org/licenses/by-sa/3.0/}{CC BY-SA
                        3.0} license.}
        \label{fig:kasner:u-param}
\end{figure}

With the exception of the null measure set $(p_l, p_m, p_n) = (0,0,1)$
and $(p_l, p_m, p_n) = (-1/3,2/3,2/3)$, Kasner indices are always
different from each other and one of them is always negative. The
$(0,0,1)$ case can be shown to be equivalent to the Minkowski
space-time. It is customary to order the Kasner indices from the
smallest to the greatest $p_1 < p_2 < p_3$. Since the three variables
$p_1, p_2, p_3$ are constrained by equations~\eqref{kasner:sum}
and~\eqref{kasner:sumofsquares}, they can be expressed as functions of
a unique parameter $u$ as
\begin{equation}\label{kasner:u-param}
p_1(u) = \tfrac{-u}{1+u+u^2},~~
p_2(u) = \tfrac{1 + u}{1+u+u^2},~~
p_3(u) = \tfrac{u(1 + u)}{1+u+u^2}
\end{equation}
\begin{equation}\label{kasner:u-range}
1 \leq u < +\infty
\end{equation}
The range and the values of the Kasner indices are portrayed in
Fig.~\ref{fig:kasner:u-param}.
Excluding the Minkowskian case $(0,0,1)$, for any choice of the Kasner
indices, the spatial metric
$\eta=\text{diag}(t^{2p_l},t^{2p_m},t^{2p_n})$ has a physical naked
singularity, called also \emph{Big Bang}, when $t = 0$ (or
$\tau\to -\infty$).

\subsection{Bianchi II}\label{sec:bianchi-ii}

A period in the evolution of the Universe when the r.h.s
of~\eqref{gener-dynam-bianchi:log-variables} can be neglected, and the
dynamics is Bianchi I-like, is called \emph{Kasner regime} or
\emph{Kasner epoch}.  In this Section we show how Bianchi II links
together two different Kasner epochs at $\tau\to -\infty$ and
$\tau\to\infty$. A series of successive Kasner epochs, where one
cosmic scale factor increases monotonically, is called a \emph{Kasner
        era}.

Again, after substituting the Bianchi II structure constants
$\lambda_l=1,\lambda_m=0,\lambda_n=0$ into
Eqs~\eqref{gener-dynam-bianchi:log-variables}, we get
\begin{subequations}\label{bianchi-ii:log-einstein}
        \begin{align}
        \label{bianchi-ii:alpha} 
        &{(q_l)}_{\tau\tau} = - e^{2q_l}\\
        \label{bianchi-ii:beta-gamma} 
        &{(q_m)}_{\tau\tau} = {(q_n)}_{\tau\tau} = e^{2q_l}
        \end{align}
\end{subequations}
It should be noted that the conditions
\begin{equation}\label{bianchi-ii:sum-relations}
{(q_l)}_{\tau\tau} + {(q_m)}_{\tau\tau} = {(q_l)}_{\tau\tau} +
{(q_n)}_{\tau\tau} = 0 
\end{equation}
hold for every $\tau$.

Let us consider the explicit solutions of
equations~\eqref{bianchi-ii:log-einstein}
\begin{subequations}\label{bianchi-ii:solution}
        \begin{align}
        {(q_l)}(\tau) = & \ln(c_1 \sech(\tau c_1 + c_2))\\
        {(q_m)}(\tau) = & 2c_3 + 2\tau c_4 - \ln(c_1 \sech(\tau c_1 + c_2))\\
        {(q_n)}(\tau) = & 2c_5 + 2\tau c_6 -  \ln(c_1 \sech(\tau c_1 + c_2))
        \end{align}
\end{subequations}
where $c_1,\dots,c_6$ are integration constants. We now analyze
how~\eqref{bianchi-ii:solution} behave as the time variable $\tau$
approaches $+\infty$ and $-\infty$ (remembering also
equation~\eqref{bianchi-i:t-tau}). By taking the asymptotic limit to
$\pm\infty$, we note that a Kasner regime at $+\infty$ is ``mapped''
to another Kasner regime at $-\infty$, but with different Kasner
indices.
\begin{equation}\label{bianchi-ii:limits}
\begin{dcases}
p_l = \tfrac{{(q_l)}_\tau}{2\Lambda} = \tfrac{-c_1 /2}{c_4 + c_6 + c_1 /2}\\
p_m = \tfrac{{(q_m)}_\tau}{2\Lambda} = \tfrac{c_4 + c_1 /2}{c_4 + c_6 + c_1 
        /2}\\
p_n = \tfrac{{(q_n)}_\tau}{2\Lambda} = \tfrac{c_6 + c_1 /2}{c_4 + c_6 + c_1 /2}
\end{dcases}~
\begin{dcases}
p'_l = \tfrac{{(q_l)}_\tau}{2\Lambda} = \tfrac{c_1 /2}{c_4 + c_6 - c_1 /2}\\
p'_m = \tfrac{{(q_m)}_\tau}{2\Lambda} = \tfrac{c_4 - c_1 /2}{c_4 + c_6 - c_1
        /2}\\
p'_n = \tfrac{{(q_n)}_\tau}{2\Lambda} = \tfrac{c_6 - c_1 /2}{c_4 +
        c_6 - c_1 /2}
\end{dcases}
\end{equation}
where we have exploited the notation of
equation~\eqref{bianchi-i:solution}. Both these sets of indices
satisfy the two Kasner relations~\eqref{kasner:sum}
and~\eqref{kasner:sumofsquares}.

The old (the unprimed ones) and the new indices (the primed ones) are
related by the BKL map, that can be given in the form of the new
primed Kasner indices $p_l',p_m',p_n'$ and $\Lambda'$ as functions of
the old ones. To calculate it we need at least four relations between the new and
old Kasner indices
$f_i(p_l,p_m,p_n,\Lambda,p'_l,p'_m,p'_n,\Lambda') = 0$ with
$i = 1,\dots,4$. Then we can invert these relations to find the BKL
map.

One $f_i$ is simply the sum of the primed Kasner indices (that is the
primed version of~\eqref{kasner:sum}). Other two relations can be
obtained from~\eqref{bianchi-ii:sum-relations}:
\begin{align}
\label{bianchi-ii:rel-1}
\Lambda(p_l + p_m) = &\Lambda'(p'_l + p'_m)\\
\label{bianchi-ii:rel-2}
\Lambda(p_l + p_n) = &\Lambda'(p'_l + p'_n)
\end{align}

The fourth relation is obtained by direct comparison of the asymptotic
limits~\eqref{bianchi-ii:limits}, and for this reason we will call it
\emph{asymptotic relation}.
\begin{equation}
\label{bianchi-ii:asympt-rel}
\Lambda p_l = - (\Lambda' p'_l)
\end{equation}
All other relations that can be obtained
from~\eqref{bianchi-ii:limits} are equivalent
to~\eqref{bianchi-ii:asympt-rel}.

Finally, by inverting the four
relations~\eqref{kasner:sum},\eqref{bianchi-ii:rel-1},
\eqref{bianchi-ii:rel-2},\\ \eqref{bianchi-ii:asympt-rel} and assuming
again that $p_l$ is the negative Kasner index $p_1$, we finally get
the classical BKL map:
\begin{gather}\label{bianchi-ii:BKL-map}
p'_l = \frac{|p_l|}{1-2|p_l|}, \quad p'_m =
\frac{p_m-2|p_l|}{1-2|p_l|},
\quad p'_n = \frac{p_n-2|p_l|}{1-2|p_l|},\\
\nonumber \Lambda' = (1-2|p_l|) \Lambda
\end{gather}
The main feature of the BKL map is the exchange of the negative index
between two different directions. Therefore, in the new epoch, the
negative power is no longer related to the $l$-direction and the
perturbation to the new Kasner regime (which is linked to the
$l$-terms on the r.h.s. of
Eq.~\eqref{gener-dynam-bianchi:log-variables}) is damped and vanishes
towards the singularity: the new (primed) Kasner regime is stable
towards the singularity.

If we then insert the parametrization~\eqref{kasner:u-param}
and~\eqref{kasner:u-range} into the BKL map~\eqref{bianchi-ii:BKL-map}
just found, we get that the map in the $u$ parameter is
simple-looking.
\begin{equation}\label{bianchi-vii:BKL-map}
\begin{dcases}
p_l = p_1(u)\\
p_m = p_2(u)\\
p_n = p_3(u)
\end{dcases}
\xrightarrow{\text{BKL map}}
\begin{dcases}
p'_l = p_2(u - 1)\\
p'_m = p_1(u - 1)\\
p'_n = p_3(u - 1)
\end{dcases}
\end{equation}
We emphasize that the role of the negative index is swapped between
the $l$ and the $m$-direction.

\subsection{Bianchi IX}\label{sec:bianchi-ix}

Here we briefly explain how the BKL map can be used to characterize
the Bianchi IX dynamics. This is possible because it can be
shown~\cite{Montani:2011} that the Bianchi IX dynamical evolution is
made up piece-wise by Bianchi I and II-like ``blocks''.

First, we take a look again at the Einstein
equations~\eqref{gener-dynam-bianchi:log-variables}: but now we
substitute the Bianchi IX structure constants
$\lambda_l=1,\lambda_m=1,\lambda_n=1$ into them to get the Einstein
equations for Bianchi IX:
\begin{equation}\label{bianchi-ix:einstein-equations}
  \begin{aligned}
    {(q_l)}_{\tau\tau} = {(b^2 - c^2)}^2 - a^4\\
    {(q_m)}_{\tau\tau} = {(a^2 - c^2)}^2 - b^4\\
    {(q_n)}_{\tau\tau} = {(a^2 - b^2)}^2 - c^4
  \end{aligned}
\end{equation}
Eq.~\eqref{gener-dynam-bianchi:constraint} stays valid for every
Bianchi model.

Again, we assume an initial Kasner regime, where the negative Kasner
index $p_1$ is associated with the $l$-direction. Thus, the
``exploding'' cosmic scale factor in the r.h.s.\ of
\eqref{bianchi-ix:einstein-equations} is identified again with $a$. By
retaining in the r.h.s.\ of \eqref{bianchi-ix:einstein-equations} only
the terms that grow towards the singularity, we obtain a system of
ordinary differential equations in time completely similar to the one
just encountered for Bianchi II~\eqref{bianchi-ii:log-einstein}, with
the only caveat that now there is no initial condition (i.e.\ no
choice of the initial Kasner indices) such that the initial or final
Kasner regimes are stable towards the singularity.

This because all the cosmic scale factors $\{a,b,c\}$ are treated on
an equal footing in the
r.h.s. of~\eqref{bianchi-ix:einstein-equations}.  After every Kasner
epoch or era, no matter on which axis the negative index is swapped,
there will always be a perturbation in the Einstein
equations~\eqref{bianchi-ix:einstein-equations} that will make the
Kasner regime unstable. We thus have an infinite series of Kasner eras
alternating and accumulating towards the singularity.

Given an initial $u$ parameter, the BKL map that takes it to the next
one $u'$ is
\begin{equation}\label{bianchi-ix:BKL-map}
  u' =
  \begin{dcases}
    u - 1 ~~ \text{for} ~~ u > 2\\
    \frac{1}{u - 1 }~~ \text{for} ~~ u \leq 2\\
  \end{dcases}
\end{equation}
Let us assume an initial $u_0 = k_0 + x_0$, where $k_0 = [u_0]$ is the
integer part and $x_0 = u_0 - [u_0]$ is the fractional part. If $x_0$
is irrational, as it is statistically and more general to infer, after
the $k_0$ Kasner epochs that make up that Kasner era, its inverse
$1/x_0$ would be irrational, too. This inversion happens infinitely
many times until the singularity is reached and is the source of the
chaotic features of the BKL
map\cite{Barrow:1981,Barrow:1982,Khalatnikov:1984hzi,Lifshitz:1971,
  Montani:1997,Szydlowski:1993}.

\section{Hamiltonian Formulation of the Mixmaster model}
\label{sec:HamiltonianFormalism}
      
In this section we summarize some important results obtained applying
the R. Arnowitt, S. Deser and C.W. Misner (ADM) Hamiltonian methods to
the Bianchi models. The main benefit of this approach is that the
resulting Hamiltonian system resembles the simple one of a pinpoint
particle moving in a potential well. Moreover it provides a method to
quantize the system through the canonical formalism.

\begin{figure}
  \includegraphics[scale=.35]{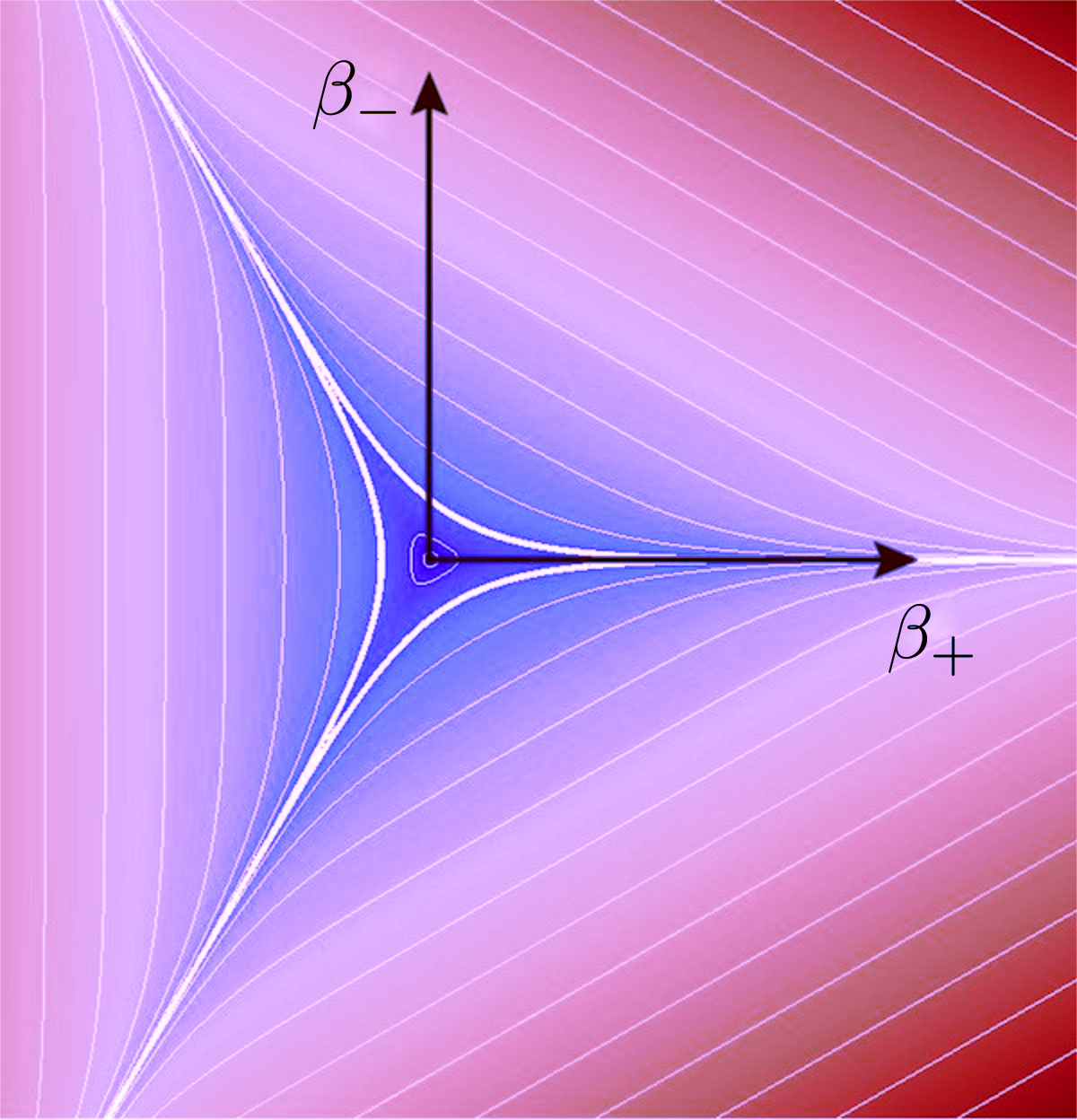}
  \caption{\label{fig:potential} Equipotentials of the
    function $V_B(\beta_\pm)$ in the $(\beta_+,\beta_-)$-plane. Here
    we can see that far from the origin the function has the symmetry
    of an equilateral triangle, while near the origin $(V_B<1)$ the
    equipotentials are closed curves.}
\end{figure}

The line element for the Bianchi IX model is:
\begin{equation}
  ds^2=N^2(t)dt^2+\frac{1}{4}e^{2\alpha}{(e^{2\beta})}_{ij}\sigma_i\sigma_j
  \label{chiara:IXLineElement}
\end{equation} 
where $N$ is the Lapse Function, characteristic of the canonical
formalism and $\sigma_i$ are 1-forms depending on the Euler angles of
the SO(3) group of symmetry. $\beta_{ij}$ is a diagonal, traceless
matrix and so it can be parameterized in terms of the two independent
variables $\beta_\pm$ as
$\beta_{ij}=(\beta_++\sqrt{3}\beta_-,\beta_+-\sqrt{3}\beta_-,-2\beta_+)$. The
factor $\frac{1}{4}$ is chosen (following \cite{Misner:1969}) so that
when $\beta_{ij}=0$, we obtain just the standard metric for a
three-sphere of radius $r=e^\alpha$. Thus for $\beta=0$ this metric is
the Robertson-Walker positive curvature metric.  The variables
$(\alpha,\beta_\pm)$ were introduced by C.W. Misner
in~\cite{Misner:1969mixmaster}; $\alpha$ describes the expansion of
the Universe, i.e. its volume changes
$(\mathcal{V}\propto e^{3\alpha})$, while $\beta_\pm$ are related to
its anisotropies (shape deformations). The kinetic term of the
Hamiltonian of the Bianchi models, written in the Misner variables,
results to be diagonal; we can then write,
following~\cite{Arnowitt:1960pr}, the super-Hamiltonian for the
Bianchi models simply as:
\begin{small}
  \begin{equation}
    \mathcal{H}_B=\frac{N\kappa}{3(8\pi)^2}e^{-3\alpha}\left(-p_\alpha^2+p_+^2+p_-^2
      +\frac{3(4\pi)^4}{\kappa^2}e^{4\alpha}V_B(\beta_\pm)\right)
    \label{chiara:Hamiltonian}
  \end{equation}
\end{small}
where $(p_\alpha,p_\pm)$ are the conjugate momenta to
$(\alpha,\beta_\pm)$ respectively and N is the lapse function.  It
should be noted that the corresponding super-Hamiltonian constraint
$\mathcal{H}_B=0$ describes the evolution of a point
$\beta=(\beta_+,\beta_-)$, that we call $\beta$-point, in function of
a new time coordinate $\alpha$ (that is the shape of the Universe in
function of its volume). Such point is subject to the
\textit{anisotropy potential} $V_B(\beta_\pm)$ which is a non-negative
function of the anisotropy coordinates, whose explicit form depends on
the Bianchi model considered. In particular Bianchi type I corresponds
to the $V_B=0$ case (free particle), Bianchi type II potential
corresponds to a single, infinite exponential wall
$\left(V_B=e^{-8\beta_+}\right)$, while for Bianchi type IX we have a
closed domain expressed by:
\begin{multline}
  V_B(\beta_\pm)=2e^{4\beta_+}\left(\cosh(4\sqrt{3}\beta_-)-1\right)+\\
  - 4e^{2\beta_+}\cosh(2\sqrt{3}\beta_-) + e^{-8\beta_+}
  \label{chiara:BianchiIXpotential}
\end{multline}
The potential ~\eqref{chiara:BianchiIXpotential} has steep exponential
walls with the symmetry of an equilateral triangle with flared open
corners, as we can see from Fig. \ref{fig:potential}. The presence of
the term $e^{4\alpha}$ in \eqref{chiara:Hamiltonian}, moreover, causes
the potential walls to move outward as we approach the cosmological
singularity $(\alpha\to-\infty)$.

Following the ADM reduction procedure, we can solve the
super-Hamiltonian constraint~\eqref{chiara:Hamiltonian} with respect
to the momentum conjugate to the new time coordinate of the
phase-space (i.e.  $\alpha$). The result is the so-called
\textit{reduced Hamiltonian} $H_{ADM}$:
\begin{equation}
  H_{ADM}\coloneqq 
  -p_\alpha=\sqrt{p_+^2+p_-^2+\frac{3(4\pi)^4}{\kappa}e^{4\alpha}V_B(\beta_\pm)}
  \label{chiara:redHamiltonian}
\end{equation}
from which we can derive the classical dynamics of the model by
solving the corresponding Hamilton's equations:
\begin{subequations}
  \begin{align}
    \beta'_\pm &\equiv \frac{d\beta_\pm}{d\alpha}= \frac{\partial 
                 H_{ADM}}{\partial p_\pm}\\
    p'_\pm &\equiv \frac{dp_\pm}{d\alpha}=-\frac{\partial H_{ADM}}{\partial 
             \beta_\pm}\\ 
    H'_{ADM} &\equiv \frac{dH_{ADM}}{d\alpha}=  \frac{\partial 
               H_{ADM}}{\partial \alpha}
  \end{align}
  \label{chiara:HamiltonEqs}
\end{subequations}
It should be noted that the choice $\dot{\alpha}=1$ fixes the temporal
gauge $N_{ADM}=\frac{6(4\pi)^2}{H_{ADM}\kappa}e^{3\alpha}$.

Since we are interested in the dynamics of the model near the
singularity $(\alpha\to-\infty)$ and because of the steepness of the
potential walls, we can consider the $\beta$-point to spend most of
its time moving as a free particle (far from the potential wall
$V_B\simeq 0$), while $V_B(\beta_\pm)$ is non-negligible only during
the short time of a bounce against one of the three walls. It follows
that we can separate the analysis into two different phases.  As far
as the free motion phase (Bianchi type I approximation) is concerned,
we can derive the velocity of the $\beta$-point from the first of Eqs.
\eqref{chiara:HamiltonEqs} (with $V_B=0$), obtaining:
\begin{equation}
  \beta'\equiv\sqrt{\beta'^2_++\beta'^2_-}=1
  \label{chiara:velocity}
\end{equation}

It remains to study the bounce against the potential. Following
\cite{Misner:1969} we can summarize the results of this analysis in
three points: (i) The position of the potential wall is defined as the
position of the equipotential in the $\beta$-plane bounding the region
in which the potential terms are significant. The wall turns out to
have velocity
$|\beta'_{wall}|\equiv\frac{d\beta_{wall}}{d\alpha}=\frac{1}{2}$, that
is one half of the $\beta$-point velocity. So a bounce is indeed
possible. (ii) Every bounce occurs according to the reflection law:
\begin{equation}
  \sin\theta_i-\sin\theta_f=\frac{1}{2}\sin(\theta_i+\theta_f)
  \label{chiara:reflectionLaw}
\end{equation} 
where $\theta_i$ and $\theta_f$ are the incidence and the reflection
angles respectively. (iii) The maximum incidence angle for the
$\beta$-point to have a bounce against a specific potential wall turns
out to be:
\begin{equation}
  \theta_{max}=\arccos\left(\frac{1}{2}\right)=\frac{\pi}{3}
  \label{chiara:maxAngle}
\end{equation}
Because of the triangular symmetry of the potential, this result
confirms the fact that a bounce against one of the three walls always
happens. It follows that the $\beta$-point undergoes an infinite
series of bounces while approaching the singularity and, sooner or
later, it will assume all the possible directions, regardless the
initial conditions.  In short we are dealing with a pinpoint particle
which undergoes an uniform rectilinear motion marked by the constants
of motion $(p_+,p_-)$ until it reach a potential wall. The bounce
against this wall has the effect of changing the values of
$p_\pm$. After that, a new phase of uniform rectilinear motion (with
different constants of motion) starts. If we remember that the free
particle case corresponds to the Bianchi type I model, whose solution
is the Kasner solution (see Sec. \ref{sec:bianchi-i}), we recover the
sequence of Kasner epochs that characterizes the BKL map
(Sec. \ref{sec:bianchi-ix}).

In conclusion, it is worth mentioning that, through the analysis of
the geometrical properties of the scheme outlined so far near the
cosmological singularity, we can obtain a sort of conservation law,
that is:
\begin{equation}
  \langle{H_{ADM}\alpha}\rangle=\text{const}
  \label{chiara:adiabaticInvariant}
\end{equation} 
Here the symbol $\langle{...}\rangle$ denotes the average value of
$H_{ADM}\alpha$ over a large number of runs and bounces. In fact,
although it is not a constant, this quantity acquires the same
constant value just before each bounce:
$H^n_{ADM}\alpha_n=H^{n+1}_{ADM}\alpha_{n+1}$. In this sense
quantities with this property will be named \textit{adiabatic
  invariants}.

\subsection{Quantization of the Mixmaster model}

Near the Cosmological Singularity some quantum effects are expected to
influence the classical dynamics of the model.  The use of the
Hamiltonian formalism enables us to quantize the cosmological theory
in the canonical way with the aim of identify such effects.

Following the canonical formalism \cite{Misner:1969}, we can introduce
the basic commutation relations
$[\hat{\beta}_a,\hat{p}_b]=i\delta_{ab}$, which can be satisfied by
choosing $\hat{p}_a=-i\frac{\partial}{\partial\beta_a}$. Hence all the
variables become operators and the super-Hamiltonian constraint
\eqref{chiara:Hamiltonian} is written in its quantum version
($\hat{\mathcal{H}}\Psi(\alpha,\beta_\pm)=0$) i.e. the Wheeler-De Witt
(WDW) equation:
\begin{small}
  \begin{equation}
    \left[\frac{\partial^2}{\partial\alpha^2}-\frac{\partial^2}{\partial\beta_+^2}-
      \frac{\partial^2}{\partial\beta_-^2}+\frac{3(4\pi)^4}{\kappa^2}e^{4\alpha}V_B(
      \beta_\pm)\right]\Psi(\alpha,\beta_\pm)=0
    \label{chiara:WDWequation}
  \end{equation}
\end{small}
The function $\Psi(\alpha,\beta_\pm)$ is the wave function of the
Universe, which we can choose of the form:
\begin{equation}
  \Psi(\alpha,\beta_\pm)=\sum_n\chi_n(\alpha)\phi_n(\alpha,\beta_\pm)
  \label{chiara:waveFunction}
\end{equation}

If we assume the validity of the \textit{adiabatic approximation} (see
\cite{Montani:2013})
\begin{equation}
  |\partial_\alpha\chi_n(\alpha)|>>|\partial_\alpha\phi_n(\alpha,\beta_\pm)|
  \label{chiara:adiabaticHypothesis}
\end{equation}
we can solve the WDW equation by separation of variables. In
particular the eigenvalues $E_n$ of the reduced Hamiltonian
\eqref{chiara:redHamiltonian} are obtained from those of the equation
$\hat{H}_{ADM}\phi_n=E_n^2\phi_n$ that is:
\begin{small}
  \begin{equation}
    \left[-\frac{\partial^2}{\partial\beta_+^2}-\frac{\partial^2}{\partial\beta_-^2}
      +\frac{3(4\pi)^4}{\kappa^2}e^{4\alpha}V_B(\beta_\pm)\right]\phi_n(\alpha,\beta_
    \pm)=E_n^2\phi_n    
  \end{equation}
\end{small}
The result, achieved by approximating the triangular potential with a
quadratic box with infinite vertical walls and the same area of the
triangular one $(A=\frac{3}{4}\sqrt{3}\alpha^2)$, turns out to be:
\begin{equation}
  E_n\sim\frac{2\pi}{3^{3/4}}\sqrt{n^2+m^2}\ \alpha^{-1}=\frac{a_n}{\alpha}
  \label{chiara:Eigenvalues}
\end{equation}
where $n,m$ are the integer and positive quantum numbers related to
the anisotropies $\beta_\pm$.

An important conclusion that C.W. Misner derived from this analysis is
that quasi-classical states (i.e. quantum states with very high
occupation numbers) are preserved during the evolution of the Universe
towards the singularity. In fact if we substitute the quasi-classical
approximation $H\simeq E_n$ in \eqref{chiara:adiabaticInvariant} and
we study it in the limit $\alpha\to-\infty$, we find:
\begin{equation}
  \langle n^2+m^2\rangle=\text{const}
\end{equation}
We can therefore conclude that if we assume that the present Universe
is in a quasi-classical state of anisotropy ($n^2+m^2>>1$), then, as
we extrapolate back towards the Cosmological Singularity, the quantum
state of the Universe remains always quasi-classical.

\section{Mixmaster model in the Polymer approach}\label{sec:mixm-poly}

Here the Hamiltonian formalism introduced in Sec.
\ref{sec:HamiltonianFormalism} is used for the analysis of the
dynamics of the modified Mixmaster model. We choose to apply the
polymer representation to the isotropic variable $\alpha$ of the
system due to its deep connection with the volume of the Universe. As
a consequence we will need to find an approximated operator for the
conjugate momentum $p_\alpha$, while the anisotropy variables
$\beta_\pm$ will remain unchanged from the standard case.

The introduction of the polymer representation consists in the formal
substitution, derived in Sec. \ref{sec:dynamics}:
\begin{equation}
  p_\alpha^2\to\frac{1}{\mu^2}\sin^2(\mu p_\alpha)
  \label{chiara:polymerSubstitution}
\end{equation}
which we apply to the super-Hamiltonian constraint $\mathcal{H}_B=0$
(with $\mathcal{H}_B$ defined in \eqref{chiara:Hamiltonian}):
\begin{multline}
  \frac{N\kappa}{3(8\pi)^2}e^{-3\alpha}\left(-\frac{1}{\mu^2}\sin^2(\mu
    p_\alpha)+p_+^2+p_-^2+\right.\\
  +\left.\frac{3(4\pi)^4}{\kappa^2}e^{4\alpha}V_B(\beta_\pm)\right)=0
  \label{chiara:polymerSuperHam}
\end{multline}
In what follows we will use the ADM reduction formalism in order to
study the new model both from a semiclassical \footnote{In this case
  the word ``semiclassical'' means that our super-Hamiltonian
  constraint was obtained as the lowest order term of a WKB expansion
  for $\hbar\to0$.} and a quantum point of view.

\subsection{Semiclassical Analysis}\label{sec:semiclassical-analysis}

First of all we derive the reduced polymer Hamiltonian\\
$(H_\alpha\coloneqq-p_\alpha)$ from \eqref{chiara:polymerSuperHam}:
\begin{equation}
  H_\alpha=\frac{1}{\mu}\arcsin\left(\sqrt{\mu^2\left[p_+^2+p_-^2+\frac{3(4\pi)^4}
        {\kappa^2}e^{4\alpha}V_B(\beta_\pm)\right]}\right)
\end{equation}
where the condition
$0\leq\mu^2(p_+^2+p_-^2+\frac{3(4\pi)^4}{\kappa^2}e^{4\alpha}V_B(\beta_\pm))\leq
1$ has to be imposed due to the presence of the function arcsine.  The
dynamics of the system is then described by the deformed Hamilton's
equations (that come from Eqs. \eqref{chiara:HamiltonEqs} with
$H_{ADM}\equiv H_\alpha$):
\begin{small}
  \begin{subequations}
    \begin{align}
      \beta'_\pm&=\frac{p_\pm}{\sqrt{(p^2+\frac{3(4\pi)^4}{\kappa^2}e^{4\alpha}V_B
                  )[1-\mu^2(p^2+\frac{3(4\pi)^4}{\kappa^2}e^{4\alpha}V_B)]}} 
                  \label{chiara:anisotropyVelocity}\\
      p'_\pm&=-\frac{\frac{3(4\pi)^4}{2\kappa^2}e^{4\alpha}\frac{\partial 
              V_B}{\partial\beta_\pm}}{\sqrt{(p^2+\frac{3(4\pi)^4}{\kappa^2}e^{4
              \alpha}V_B)[1-\mu^2(p^2+\frac{3(4\pi)^4}{\kappa^2}e^{4\alpha}V_B
              )]}}\\
      H'_\alpha&=\frac{\frac{6(4\pi)^4}{\kappa^2}e^{4\alpha}V_B}{\sqrt{[1-\mu^2(p^
                 2+\frac{3(4\pi)^4}{\kappa^2}e^{4\alpha}V_B)](p^2+\frac{3(4\pi)^4}{
                 \kappa^2}e^{4\alpha}V_B)}}
    \end{align}
    \label{chiara:polymerHamEq}
  \end{subequations}
\end{small}
where the prime stands for the derivative with respect to $\alpha$ and
$p^2=p_+^2+p_-^2$.

\begin{figure}
  \includegraphics[scale=.6]{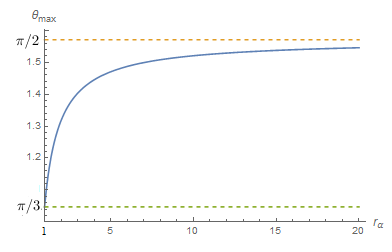}
  \caption{\label{fig:polymerMaxAngle} The maximum
    angle to have a bounce as a function of the anisotropy velocity
    $r_\alpha\in[1,\infty)$. The two limit values
    $\frac{\pi}{3}\ (r_\alpha=1)$ and
    $\frac{\pi}{2}\ (r_\alpha\to\infty)$ are highlited.}
\end{figure}

As for the standard case, we start by studying the simplest case of
$V_B=0$ (Bianchi I approximation), which corresponds to the phase when
the $\beta$-point is far enough from the potential walls. Here the
persistence of a cosmological singularity can be easily demonstrated
since $\alpha$ results to be linked to the time variable $t$ through a
logarithmic relation (as we will see later in
Eq. \eqref{polymer-bianchi-i:solution-alpha}):
$\alpha\sim\ln(t)\underset{t\to0}{\longrightarrow}-\infty$.  Moreover
the anisotropy velocity of the particle can be derived from Eq.
\eqref{chiara:anisotropyVelocity} (while $p_\pm$ and $H_\alpha$ are
constants of motion) and the result is:
\begin{equation}
  \beta'\equiv\sqrt{\beta'^2_++\beta'^2_-}=\frac{1}{\sqrt{1-\mu^2p^2}}\coloneqq 
  r_\alpha(\mu, p_\pm)
  \label{chiara:polymerAniVel}
\end{equation}
Thus the modified anisotropy velocity depends on the constants of
motion of the system, and it turns out to be always greater than the
standard one $(r_\alpha(\mu, p_\pm)>1,\ \forall
p_\pm\in\mathbb{R})$. On the other side the velocity of each potential
wall is the same of the standard case
$\left(|\beta'_{wall}|=\frac{1}{2}\right)$. In fact the Bianchi II
potential (i.e. our approximation of a single wall of the Bianchi IX
potential) can be written in terms of the determinant of the metric
$(\sqrt{\eta}=e^{3\alpha})$ as
\begin{equation}
  e^{4\alpha-8\beta_+}=(\sqrt{\eta})^{\frac{4}{3}-\frac{8\beta_+}{3\alpha}}
  \label{chiara:BianchiIIpotential}
\end{equation}
which, in the limit $\alpha\to-\infty$, tends to the
$\Theta$-function:
\begin{equation}
  e^{4\alpha}V_B=\begin{cases}
    \infty\quad &\text{if}\quad \frac{4}{3}-\frac{8\beta_+}{3\alpha}<0\\
    0\quad &\text{if}\quad \frac{4}{3}-\frac{8\beta_+}{3\alpha}>0\\
  \end{cases}
\end{equation}
It follows that the position of the wall is defined by the equation
$\frac{4}{3}-\frac{8\beta_+}{3\alpha}=0$, i.e.
$|\beta_{wall}|=\frac{\alpha}{2}$.  The $\beta$-particle is hence
faster than the potential, therefore a bounce is always possible also
after the polymer deformation. For this reason, and from the fact that
the singularity is not removed, we can state that there will certainly
be infinite bounces onto the potential walls.  Moreover, the more the
point is moving fast with respect to the walls, the more the bounces
frequency is high.  All these facts, are hinting at the possibility
that chaos is still present in the Bianchi IX model in our Polymer
approach. It is worth noting that this is a quite interesting result if
we consider the strong link between the Polymer and the Loop Quantum
Cosmology, which tends to remove the chaos and the cosmological
singularity \cite{Ashtekar:2006es,Ashtekar:2009vc}.  In particular,
when dealing with PQM, a spatial lattice of spacing $\mu$ must be
introduced to regularize the theory
(Sec~\ref{sec:continuumLimit}). Hence, one would naively think that
the Universe cannot shrink past a certain threshold volume
(singularity removal). Actually we know that Loop Quantum Cosmology
applied both to FRW \cite{Ashtekar:2006es} and Bianchi I
\cite{Ashtekar:2009vc} predicts this \textit{Big Bounce}-like dynamics
for the primordial Universe. Anyway the persistence of the
cosmological singularity in our model seems to be a consequence of our
choice of the configurational variables, while its chaotic dynamics
towards such a singularity is expected to be a more roboust physical
feature.  In order to analyze a single bounce we need to parameterize
the anisotropy velocity in terms of the incident and reflection angles
($\theta_i$ and $\theta_f$, shown in Fig. \ref{Bounce2}):
\begin{equation}
  \begin{array}{cc}
    (\beta'_-)_i=r_{\alpha_i}\sin\theta_i  & 
                                             (\beta'_+)_i=-r_{\alpha_i}\cos\theta_i\\
    (\beta'_-)_f=r_{\alpha_f}\sin\theta_f & (\beta'_+)_f=r_{\alpha_f}\cos\theta_f
  \end{array}
  \label{chiara:velocityParameterization}
\end{equation}
As usual the subscripts $i$ and $f$ distinguish the initial quantities
(just before the bounce) from the final ones (just after the bounce).
Now we can derive the maximum incident angle for having a bounce
against a given potential wall. If we consider, for example, the left
wall of Fig.  \ref{fig:potential} we can observe that a bounce occurs
only when $(\beta'_+)_i>\beta'_{wall}=\frac{1}{2}$, that is:
\begin{equation}
  \theta_{max}^\alpha=\arccos\left(\frac{1}{2r_{\alpha_i}}\right)\simeq\frac{\pi}{
    3}+\frac{1}{2\sqrt{3}}\mu^2p^2
  \label{chiara:polyMaxAngle}
\end{equation}
Two important features should be underlined about this result: the
first one is that $\theta_{max}^\alpha$ is bigger than the standard
maximum angle $\left(\theta_{max}=\frac{\pi}{3}\right)$, as we can
immediately see from its second order expansion. The second one is
that when the anisotropy velocity tends to infinity one gets
$\theta_{max}^\alpha\to\frac{\pi}{2}$ (Fig.
\ref{fig:polymerMaxAngle}). Both these characteristics along with the
triangular symmetry of the potential confirm the presence of an
infinite series of bounces.
Each bounce involves new values for the constants of motion of the
free particle ($p_\pm, H_{\alpha}$) and a change in its direction. The
relation between the directions before and after a bounce can be
inferred by considering the Hamilton's equations
\eqref{chiara:polymerHamEq} again, this time with $V_B=e^{-8\beta_+}$.

First of all we observe that $V_B$ depends only from $\beta_+$, from
which we deduce that $p_-$ is yet a constant of motion. A second
constant of motion is $K\coloneqq H_\alpha-\frac{1}{2}p_+$, as we can
verify from the Hamilton equations \eqref{chiara:polymerHamEq}.  An
expression for $p_\pm$ can be derived from Eq.
\eqref{chiara:anisotropyVelocity}:
\begin{equation}
  p_\pm=\frac{1}{\mu}\beta'_\pm\sin(\mu H_\alpha)\sqrt{1-\sin^2(\mu H_\alpha)}
\end{equation}
where the compact notation
$\sqrt{\mu^2(p^2+\frac{3(4\pi)^4}{\kappa^2}e^{4\alpha-8\beta_+})}=\\ \sin(\mu
H_\alpha)$ is used. By the substitution of this expression and of the
parameterization \eqref{chiara:velocityParameterization} in the two
constants of motion we obtain two modified conservation laws:
\begin{subequations}
  \begin{multline}
    r_{\alpha_i}\sin(\theta_i)\sin(\mu H_i)\sqrt{1-\sin^2(\mu H_i)}=\\
    =r_{\alpha_f}\sin(\theta_f)\sin(\mu H_f)\sqrt{1-\sin^2(\mu H_f)}
    \label{chiara:ConsModP-}
  \end{multline}
  \begin{multline}
    H_i+\frac{1}{2 \mu}r_{\alpha_i}\cos(\theta_i)\sin(\mu
    H_i)\sqrt{1-\sin^2(\mu H_i)}=\\
    =H_f-\frac{1}{2 \mu}r_{\alpha_f}\cos(\theta_f)\sin(\mu
    H_f)\sqrt{1-\sin^2(\mu H_f)}
    \label{chiara:ConsModK}
  \end{multline}
  \label{chiara:ConsMod}
\end{subequations}
The reflection law can be then derived after a series of algebraic
passages which involve the substitution of \eqref{chiara:ConsModP-} in
\eqref{chiara:ConsModK}, and the use of the explicit expression
\eqref{chiara:polymerAniVel} for the anisotropy velocity $r_\alpha$:
\begin{multline}
  \frac{1}{2} \mu p_f \left(\cos\theta_f+\frac{\cos \theta_i
      \sin\theta_f}{\sin\theta_i}\right)=\\
  =\arcsin(\mu p_f)-\arcsin(\mu p_i)
  \label{chiara:reflLaw}
\end{multline}
In order to have this law in a simpler form, comparable to the
standard case, an expansion up to the second order in $\mu p<<1 $ is
required. The final result is:
\begin{equation}
  \frac{1}{2}\sin(\theta_i+\theta_f)=\sin\theta_i(1+\Pi_f^2)-\sin\theta_f(1+\Pi_i^
  2)
  \label{chiara:secondOrderReflLaw}
\end{equation}
where we have defined $\Pi^2=\frac{1}{6}\mu^2p^2$.

\subsection{Comparison with previous models}\label{sec:comparison}

As we stressed in Sec. \ref{sec:semiclassical-analysis} our model
turns out to be very different from Loop Quantum Cosmology
\cite{Ashtekar:2006es,Ashtekar:2009vc}, which predicts a Big
Bounce-like dynamics. Moreover, if the polymer approach is applied to
the anisotropies $\beta_\pm$, leaving the volume variable $\alpha$
classical \cite{Montani:2013}, a singular but non-chaotic dynamics
is recovered.

It should be noted that the polymer approach in the perturbative limit
can be interpreted as a modified commutation relation
\cite{Battisti:2008du}:
\begin{equation} [\hat{q},\hat{p}]=i(1-\mu p^2)
\end{equation}
where $\mu>0$ is the deformation parameter. Another quantization
method involving an analogue modified commutation relation is the one
deriving by the Generalized Uncertainty Principle (GUP):
\begin{equation}
  \Delta q\Delta p\geq\frac{1}{2}(1+s(\Delta p)^2+s\braket{p}^2),\quad (s>0)
  \label{math:GUP}
\end{equation}
through which a minimal length is introduced and which is linked to
the commutation relation:
\begin{equation} [\hat{q},\hat{p}]=i(1+sp^2)
\end{equation}
In \cite{BattistiGUP:2009} the Mixmaster model in the GUP approach
(applied to the anisotropy variables $\beta_\pm$) is developed and the
resulting dynamics is very different from the one deriving by the
application of the polymer representation to the same variables
\cite{Montani:2013}.  The GUP Bianchi I anisotropy velocity turns out
to be always greater than $1$ ($\beta_{GUP}'^2=1+6sp^2+9s^2p^4$) and,
in particular, greater than the wall velocity $\beta'^{GUP}_{wall}$
(which in this case is different from $\frac{1}{2}$). Moreover the
maximum angle of incidence for having a bounce is always greater than
$\frac{\pi}{3}$. Therefore the occurrence of a bounce between the
$\beta$-particle and the potential walls of Bianchi IX can be
deduced. The GUP Bianchi II model is not analytic (no reflection law
can be inferred), but some qualitative arguments lead to the
conclusion that the deformed Mixmaster model can be considered a
chaotic system.  In conclusion if we apply the two modified
commutation relations to the anisotropy (physical) variables of the
system we recover very different behaviors. Instead the polymer
applied to the geometrical variable $\alpha$ leads to a dynamics very
similar to the GUP one described here: the anisotropy velocity
\eqref{chiara:polymerAniVel} is always greater than $1$ and the
maximum incident angle \eqref{chiara:polyMaxAngle} varies in the range
$\frac{\pi}{3}<\theta_{max}<\frac{\pi}{2}$, so that the chaotic
features of the two models can be compared.

\subsection{Quantum Analysis}

The modified super-Hamiltonian constraint
\eqref{chiara:polymerSuperHam} can be easily quantized by promoting
all the variables to operators and by approximating $p_\alpha$ through
the substitution \eqref{chiara:polymerSubstitution}:
\begin{multline}
  \left[-\frac{1}{\mu^2}\sin^2(\mu p_\alpha)+\hat{p}_+^2+\hat{p}_-^2+\right.\\
  +\left.\frac{3(4\pi)^4}{\kappa^2}e^{4\alpha}V_B(\hat{\beta}_\pm)\right]\psi(p_
  \alpha,p_\pm)=0
  \label{chiara:polymerQuantumEq}
\end{multline}
If we consider the potential walls as perfectly vertical, the motion
of the quantum $\beta$-particle boils down to the motion of a free
quantum particle with appropriate boundary conditions.  The free
motion solution is obtained by imposing $V_B=0$ as usual, and by
separating the variables in the wave function:
$\psi(p_\alpha,p_\pm)=\chi(p_\alpha)\phi(p_\pm)$.  It should be noted
that, although we can no longer take advantage of the adiabatic
approximation \eqref{chiara:adiabaticHypothesis}, due to the necessary
use of the momentum polarization in Eq.
\eqref{chiara:polymerQuantumEq}, nonetheless we can suppose that the
separation of variables is reasonable. In fact in the limit where the
polymer correction vanishes ($\mu\to0$), the Misner solution (for
which the adiabatic hypothesis is true) is recovered.

The anisotropy component $\phi(p_\pm)$ is obtained by solving the
eigenvalues problem:
\begin{equation}
  (\hat{p}_+^2+\hat{p}_-^2)\phi(p_\pm)=k^2\phi(p_\pm)
\end{equation}
where $k^2=k_+^2+k_-^2$ ($k_\pm$ are the eigenvalues of $\hat{p}_\pm$
respectively). Therefore the eigenfunctions are
\\$\phi(p_\pm)=\phi_+(p_+)\phi_-(p_-)$ with:
\begin{subequations}
  \begin{align}
    \phi_+(p_+)=A\delta(p_+-k_+)+B\delta(p_++k_+)\\
    \phi_-(p_-)=C\delta(p_--k_-)+D\delta(p_-+k_-)
  \end{align}
  \label{chiara:polyEigenfunctions}
\end{subequations}
where $A,B,C,D$ are integration constants. By substituting this result
in \eqref{chiara:polymerQuantumEq} we find also the isotropic
component $\chi(p_\alpha)$ and the related eigenvalue
$\bar{p}_\alpha$:
\begin{align}
  \chi(p_\alpha)=\delta(p_\alpha-\bar{p}_\alpha)\\
  \bar{p}_\alpha=\frac{1}{\mu}\arcsin(\mu k)
\end{align}

The potential term is approximated by a square box with vertical
walls, whose length is expressed by $L(\alpha)=L_0+\alpha$, so that
they move outward with velocity $|\beta'_{wall}|=\frac{1}{2}$:
\begin{equation}
  V_B(\alpha,\beta_\pm)=\begin{cases}
    0 \quad\ \text{if}\quad 
    -\frac{L(\alpha)}{2}\leq\beta_\pm\leq\frac{L(\alpha)}{2}\\
    \infty\quad\text{elsewhere}
  \end{cases}
\end{equation}
The system is then solved by applying the boundary conditions:
\begin{equation}
  \begin{cases}
    \phi_+\left(\frac{L(\alpha)}{2}\right)=\phi_+\left(-\frac{L(\alpha)}{2}\right)=0
    \\
    \phi_-\left(\frac{L(\alpha)}{2}\right)=\phi_-\left(-\frac{L(\alpha)}{2}\right)=0
  \end{cases}
\end{equation}
to the eigenfunctions \eqref{chiara:polyEigenfunctions}, expressed in
coordinate representation (i.e. $\phi(\beta_\pm)$, which can be
calculated through a standard Fourier transformation).  The final
solution is:
\begin{multline}
  \phi_{n,m}(\beta_\pm)=\frac{1}{2L(\alpha)} \left( e^{ik^+_n
      \beta_+}-e^{-i
      k^+_n \beta_+}e^{-i n \pi}\right)\cdot\\
  \cdot \left( e^{ik^-_m \beta_-}-e^{-ik^-_m \beta_-}e^{-i m
      \pi}\right)
  \label{chiara:anisoPolymerWaveFun}
\end{multline}
with:
\begin{equation}
  k^+_n=\frac{n \pi}{L(\alpha)}; \qquad k^-_m=\frac{m \pi}{L(\alpha)} 
\end{equation}
Instead the isotropic component results to be:
\begin{align}
  \bar{p}_\alpha=\frac{1}{\mu}\arcsin\left(\frac{\mu\pi}{L(\alpha)}\sqrt{m^2+n^2}
  \right)\label{chiara:polyEigenvalues}\\
  \chi(\alpha)=e^{\frac{i}{\mu}\underset{0}{\overset{x}{\int }}dt 
  \arcsin\left(\frac{\mu\pi}{L(\alpha)}\sqrt{m^2+n^2}\right)}
  \label{chiara:isoPolymerWaveFun}
\end{align}
We emphasize the fact that, as we announced at the beginning of this
section, the Misner solution \eqref{chiara:Eigenvalues} can be easily
recovered from Eq.  \eqref{chiara:polyEigenvalues}, by taking the
limit $\mu\to0$.

\subsection{Adiabatic Invariant}

\begin{figure}
  \includegraphics[scale=.5]{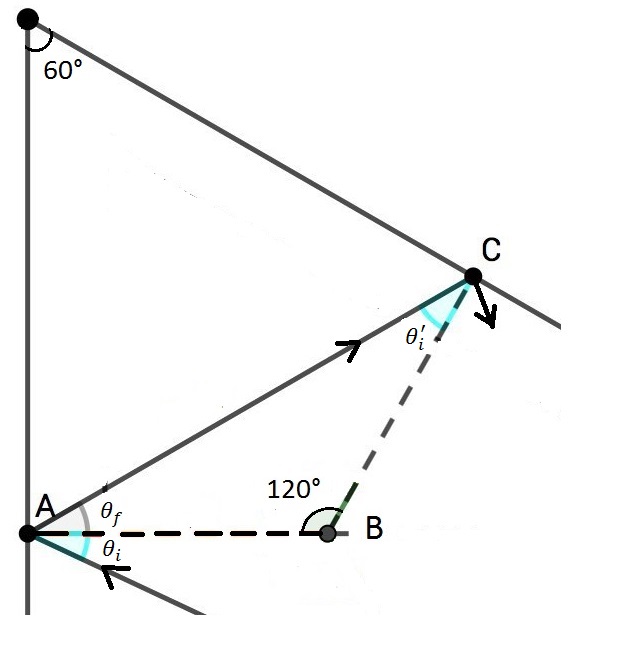}
  \caption{Relation between the angles of incidence of
    two following bounces: $\theta_i$ and $\theta'_i$. Here we can see
    that the condition $\theta_f+\theta'_i=60^\circ$ must hold in
    order to have $180^\circ$ as the sum of internal angles of the
    triangle $A\overset{\triangle}{B}C$.  It follows that the initial
    condition $\theta_i+\theta_f=60^\circ$ implies
    $\theta_i=\theta_i'$ for every couple of subsequent bounces.}
  \label{Bounce2}
\end{figure}

\begin{figure}
  \includegraphics[scale=.35]{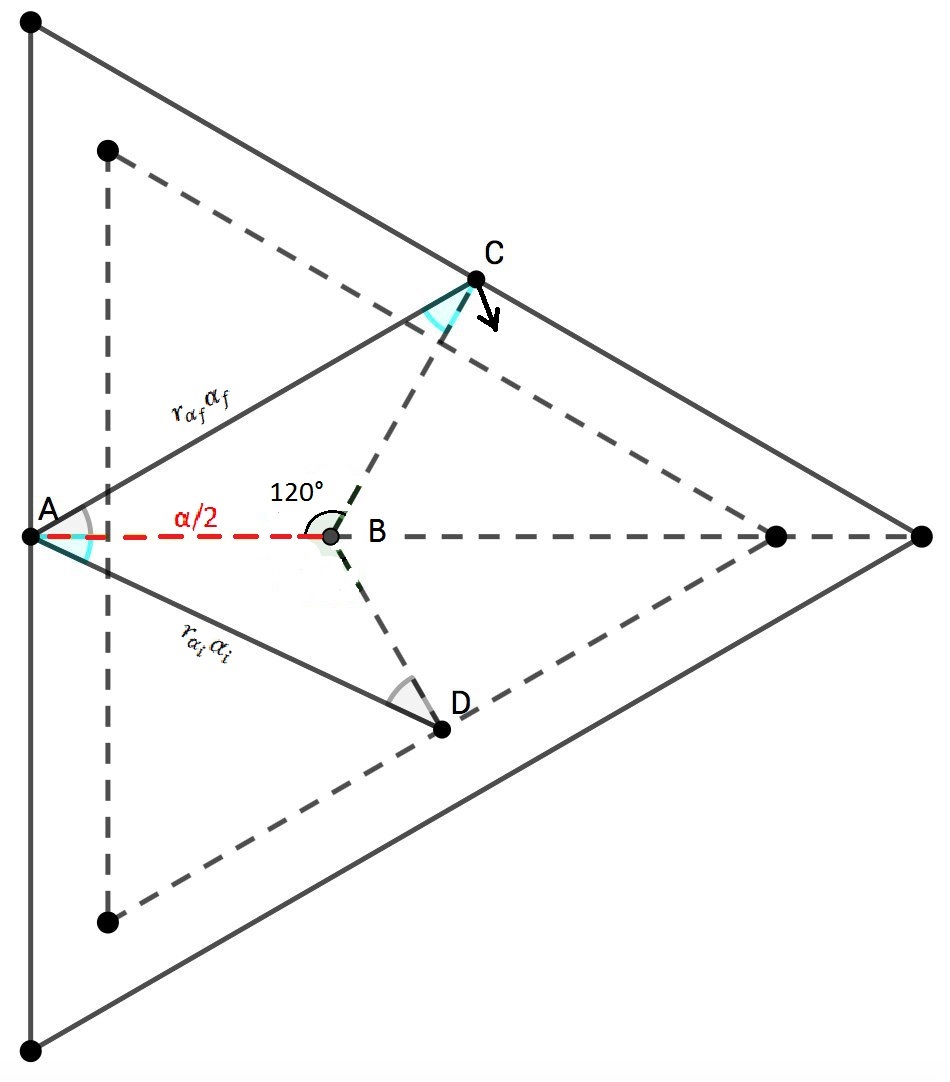}
  \caption{Geometric relations between two successive
    bounces.}
  \label{Bounce}
\end{figure}

Because of the various hints about the presence of a chaotic behavior,
outlined in Sec. \ref{sec:semiclassical-analysis} and that will be
confirmed by the analysis of the Poincar\'e map in Sec
\ref{sec:polymer-BKL-map}, we can reproduce a calculation similar to
the Misner's one \cite{Misner:1969} in order to obtain information
about the quasi-classical properties of the early universe. In Sec.
\ref{sec:semiclassical-analysis} we found that the $\beta$-point
undergoes an infinite series of bounces against the three potential
walls, moving with constant velocity between a bounce and
another. Every bounce implies a decreasing of $H_\alpha$ due to the
conservation of $K=H_\alpha-\frac{1}{2}p_+$ and the definition of a
new direction of motion through Eq.
\eqref{chiara:secondOrderReflLaw}. Because of the ergodic properties
of the motion we can assume that it won't be influenced by the choice
of the initial conditions. Thus we can choose them in order to
simplify the subsequent analysis. A particularly convenient choice is:
$\theta_i+\theta_f=60^\circ$ for the first bounce. In fact as a
consequence of the geometrical properties of the system all the
subsequent bounces will be characterized by the same angle of
incidence: $\theta'_i=\theta_i$ (Fig. \ref{Bounce2}).
Let us now analyze a single bounce, keeping in mind Fig. \ref{Bounce}.
At the time $\alpha$ in which we suppose the bounce to occur, the wall
is in the position $\beta_{wall}=\frac{|\alpha|}{2}$. Moving between
two walls, the particle has the constant velocity $|\beta'|=r_\alpha$
(Eq.  \eqref{chiara:polymerAniVel}), so that we can deduce the spatial
distance between two bounces in terms of the ``time'' $|\alpha|$
necessary to cover such a distance:
\begin{subequations}
  \begin{align}
    [\text{distance covered before the bounce}]=r_{\alpha_i}|\alpha_i|\\
    [\text{distance covered after the bounce}]=r_{\alpha_f}|\alpha_f|
  \end{align}
\end{subequations}
Now if we apply the Sine theorem to the triangles
$A\overset{\triangle}{B}C$ and $A\overset{\triangle}{B}D$, we find the
relation:
\begin{equation}
  \frac{r_{\alpha_i}|\alpha_i|}{r_{\alpha_f}|\alpha_f|}=\frac{\sin\theta_i}{\sin
    \theta_f}
  \label{chiara:constAlpha1}
\end{equation}
Moreover, from \eqref{chiara:ConsModP-} one gets:
\begin{equation}
  \frac{r_{\alpha_f}\sin(\mu H_{\alpha_f})\sqrt{1-\sin^2(\mu 
      H_{\alpha_f})}}{r_{\alpha_i}\sin(\mu H_{\alpha_i})\sqrt{1-\sin^2(\mu 
      H_{\alpha_i})}}=\frac{\sin\theta_i}{\sin\theta_f}=\text{cost}
  \label{chiara:constAlpha2}
\end{equation}
From these two relations we can then deduce a quantity which remains
unaltered after the bounce:
\begin{multline}
  r_{\alpha_i}^2 |\alpha_i| \sin(\mu H_i) \sqrt{1- \sin^2(\mu H_i)}=\\
  =r_{\alpha_f}^2 |\alpha_f| \sin(\mu H_f) \sqrt{1- \sin^2(\mu H_f)}
  \label{chiara:singleBounce}
\end{multline}
Now we can generalize the argument to the n-th bounce (taking place at
the ``time'' $\alpha_n$ instead of $\alpha$). By using again the Sine
theorem (applied to the triangle $A\overset{\triangle}{B}D$) we get
the relation
$r_{\alpha_i}|\alpha_i|=\frac{\sin120^\circ}{2\sin\theta_f}|\alpha_n|=C|\alpha_n
|$, where $C$ takes the same value for every bounce. Correspondingly
$r_{\alpha_f}|\alpha_f|=C|\alpha_{n+1}|$. If we substitute these new
relations in Eq. \eqref{chiara:singleBounce},we can then conclude
that:
\begin{equation}
  \langle r_\alpha |\alpha| \sin(\mu H) \sqrt{1-\sin^2(\mu H)} \rangle=\text{cost}
  \label{chiara:polymerAdiabaticInv}
\end{equation} 
where the average value is taken over a large number of bounces.  As
we are interested in the behavior of quantum states with high
occupation number, remembering that for such states the
quasi-classical approximation $H_\alpha\simeq\bar{p}_\alpha$ is valid,
we can substitute Eq.  \eqref{chiara:polyEigenvalues} and
Eq. \eqref{chiara:polymerAniVel} in \eqref{chiara:polymerAdiabaticInv}
and evaluate the correspondent quantity in the limit
$\alpha\to-\infty$. The result is that also in this case the
occupation number is an adiabatic invariant:
\begin{equation}
  \langle \sqrt{m^2+n^2} \rangle = \text{cost}
\end{equation}
that is the same conclusion obtained by Misner without the polymer
deformation.  It follows that also in this case a quasi-classical
state of the Universe is allowed up to the initial singularity.

\section{Semiclassical solution for the Bianchi I and II models}
\label{sec:polymer-bkl-map}

As we have already outlined in Sec.~\ref{sec:mixm-model:-class}, to
calculate the BKL map it is necessary first to solve the Einstein's
Eqs for the Bianchi I and Bianchi II models.

Therefore, in Sec.~\ref{sec:polymer-bianchi-i-dynamics} we solve the
Hamilton's Eqs in Misner variables for the Polymer Bianchi I, in the
semiclassical approximation introduced in
Sec.~\ref{sec:semiclassical-analysis}.

Then in Sec.~\ref{sec:parametrization} we derive a parametrization for
the Polymer Kasner indices. It is not possible anymore to parametrize
the three Kasner indices using only one parameter $u$ as
in~\eqref{kasner:u-param}, but two parameters are needed. This is due
to a modification to the Kasner
constraint~\eqref{kasner:sumofsquares}.

In Sec.~\ref{sec:polymer-bianchi-ii} we calculate how the Bianchi II
Einstein's Eqs are modified when the PQM is taken into account. Then
we derive a solution for these Eqs, using the lowest order
perturbation theory in $\mu$.  We are therefore assuming that $\mu$ is
little with respect to the Universe volume cubic root.

\subsection{Polymer Bianchi I}
\label{sec:polymer-bianchi-i-dynamics}

As already shown in Sec.~\ref{sec:semiclassical-analysis}, the Polymer
Bianchi I Hamiltonian is obtained by simply setting $V_B = 0$
in~\eqref{chiara:polymerSuperHam}. Upon solving the Hamilton's Eqs
derived from the Hamiltonian~\eqref{chiara:polymerSuperHam} in the
case of Bianchi I, variables $p_\alpha$ and $p_\pm$ are recognized to
be constants throughout the evolution. We then choose the time gauge
so as to have a linear dependence of the volume $\mathcal{V}$ from the
time $t$ with unity slope, as in the standard
solution~\eqref{bianchi-i:solution}:
\begin{equation}\label{polymer-bianchi-i:time-gauge}
N = - \frac{64 \pi^2 \mu}{\kappa \sin(2{\mu} p_\alpha)}
\end{equation}
Moreover, to have a Universe expanding towards positive times, we must
restrict the range of $\sin(2{\mu} p_\alpha) < 0$. This condition
reflects on $p_\alpha$ as
\begin{equation}\label{polymer-bianchi-i:p-alpha-range}
-\frac{\pi}{2\mu} + \frac{n\pi}{\mu} < p_\alpha <
\frac{n\pi}{\mu},\qquad p_\alpha \in \Z
\end{equation}
The branch connected to the origin $n=0$, will be called
\emph{connected-branch}. For $\mu \rightarrow 0$, it gives the correct
continuum limit: $p_\alpha < 0$. Even though, for different choices of
$n\neq0$, we cannot provide a simple and clear explanation of their
physical meaning, when we will define the parametrization of the
Kasner indices in Sec.~\ref{sec:parametrization}, we will be forced to
consider another branch too, in addition to the connected-one. We
choose arbitrarily the $n=1$ branch and, by analogy, we will call it
the \emph{disconnected-branch}.

The solution to Hamilton's Eqs for Polymer Bianchi I in
the~\eqref{polymer-bianchi-i:time-gauge} time gauge, reads as
\begin{subequations}\label{polymer-bianchi-i:solution}
        \begin{align}
        \label{polymer-bianchi-i:solution-alpha}
        \alpha(t) = & ~\frac{1}{3} \ln(t)\\
        \label{polymer-bianchi-i:solution-beta}
        \beta_\pm(t) = & - \frac{2}{3} \frac{\mu p_\pm}{\sin(2{\mu}
                p_\alpha)} \ln(t) = 
        - \frac{2\mu p_\pm}{\sin(2{\mu} p_\alpha)}
        \alpha(t)
        \end{align}
\end{subequations}
Since $p_\alpha$, $p_\pm$ and $\mu$ in
the~\eqref{polymer-bianchi-i:solution} are all numerical constants,
it is qualitatively the very
same as the classical solution~\eqref{bianchi-i:solution}.
In particular, even in the Polymer case the volume
$\mathcal{V}(t) \propto e^{3\alpha(t)} = t$ goes to zero when
$t \rightarrow 0$: i.e.\ \emph{the singularity is not removed} and
we have a Big Bang, as already pointed out in
Sec.~\ref{sec:semiclassical-analysis}.

By making a reverse canonical transformation from Misner variables
($\alpha$, $\beta_\pm$) to ordinary ones ($q_l$, $q_m$, $q_n$), we
find out that~\eqref{kasner:sum} holds unchanged
while~\eqref{kasner:sumofsquares} is modified into:
\begin{equation}\label{polymer-bianchi-i:sumofsquares-p}
{p_l}^2 + {p_m}^2 + {p_n}^2 = \frac{1}{3} \left( 1 +
\frac{2}{\cos^2(\mu p_\alpha)}\right)
\end{equation}

If the continuum limit $\mu \rightarrow 0$ is taken, we get back the
old constraint~\eqref{kasner:sumofsquares} as expected. Since the relation
\begin{equation}\label{kasner-vs-anisotropy-velocity}
	{p_l}^2 + {p_m}^2 + {p_n}^2 = \frac{1}{3}+\frac{2}{3}\beta'^2	
\end{equation}
holds, the inequality
${p_l}^2 + {p_m}^2 + {p_n}^2 \geq 1,\quad \forall p_\alpha \in \R$ is
directly linked to $\beta$-point velocity~\eqref{chiara:polymerAniVel}
being always equal or greater than one, as demonstrated in
Sec.~\ref{sec:semiclassical-analysis}. This is a noteworthy difference
with respect to the standard case of Eq.~\eqref{chiara:velocity},
where the speed was always one.

We can derive an alternative expression for the Kasner
constraint~\eqref{polymer-bianchi-i:sumofsquares-p} by exploiting the
notation of Eqs~\eqref{bianchi-i:solution}:
\begin{equation}\label{polymer-bianchi-i:sumofsquares-exact}
{p_l}^2 + {p_m}^2 + {p_n}^2 = \frac{1}{3} \left( 1 + 
\frac{4}{1 \pm \sqrt{1 - Q^2 }} \right)
\end{equation}
where the plus sign is for the connected-branch while the minus for
the disconnected-branch. We have defined the dimensionless quantity
$Q$ as
\begin{equation}\label{polymer-bianchi-i:Q}
Q := \frac{{(8\pi)}^2 \mu \Lambda }{\kappa }
\end{equation}
It is worth noticing that the lattice spacing $\mu$ has been
incorporated in $Q$, so that the continuum limit $\mu\to 0$ is now
equivalently reached when $Q\to 0$.
Condition~\eqref{polymer-bianchi-i:p-alpha-range} reflects on the
allowed range for Q:\
\begin{equation}\label{polymer-bianchi-i:Q-range}
\abs*{Q} \leq 1
\end{equation}
both for the connected and disconnected branches.

\subsection{Parametrization of the Kasner indices}\label{sec:parametrization}

In this Section we define a parametrization of the Polymer Kasner
indices.  In the standard case of~\eqref{kasner:u-param}, because
there are three Kasner indices and two Kasner
constraints~\eqref{kasner:sum} and~\eqref{kasner:sumofsquares}, only
one parameter $u$ is needed to parametrize the Kasner indices. On the
other hand, in the Polymer case, even if there are two Kasner
constraints~\eqref{kasner:sum}
and~\eqref{polymer-bianchi-i:sumofsquares-exact} as well,
constraint~\eqref{polymer-bianchi-i:sumofsquares-exact} already
depends on another parameter $Q$. This means that any parametrization
of the Polymer Kasner indices will inevitably depend on two
parameters, that we arbitrarily choose as $u$ and $Q$, where $u$ is
defined on the same range as the standard case~\eqref{kasner:u-param}
and $Q$ was defined in~\eqref{polymer-bianchi-i:Q}. They are both
dimensionless. We will refer to the following expressions as the
$(u,Q)$-parametrization for the Polymer Kasner indices:
\begin{align}\label{parametrization:u-Q-param}
\begin{split}
p_1 & = \frac{1}{3}\left\{ 1 - \frac{1 + 4u + u^2}{1 + u + u^2}
{\left\lbrack \half \left( 1 \pm \sqrt{1 - Q^2} \right)
        \right\rbrack}^{-\half} \right\}\\
p_2 & = \frac{1}{3}\left\{ 1 + \frac{2 + 2u - u^2}{1 + u + u^2}
{\left\lbrack \half \left( 1 \pm \sqrt{1 - Q^2} \right)
        \right\rbrack}^{-\half} \right\}\\
p_3 & = \frac{1}{3}\left\{ 1 +\frac{-1 + 2u + 2u^2}{1 + u + u^2}
{\left\lbrack \half \left( 1 \pm \sqrt{1 - Q^2} \right)
        \right\rbrack}^{-\half} \right\}
\end{split}
\end{align}
where the plus sign is for the connected branch and the minus sign is
for the disconnected branch. By construction, the standard
$u$-parametrization~\eqref{kasner:u-param} is recovered in the limit
$Q \rightarrow 0$ of the connected branch. We will see clearly in
Sec.~\ref{sec:BKL-map-properties} how the $Q$ parameter can be thought
of as a measure of the \emph{quantization degree} of the Universe. The
more the $Q$ of the connected-branch is big in absolute value, the
more the deviations from the standard dynamics due to PQM are
pronounced. The opposite is true for the disconnected-branch.  To gain
insight on multiple-parameters parametrizations of the Kasner indices,
the reader can refer to\cite{Montani:2004mq,Elskens:1987}.

\begin{figure}
        \begin{center}
                \includegraphics[width=1\columnwidth]{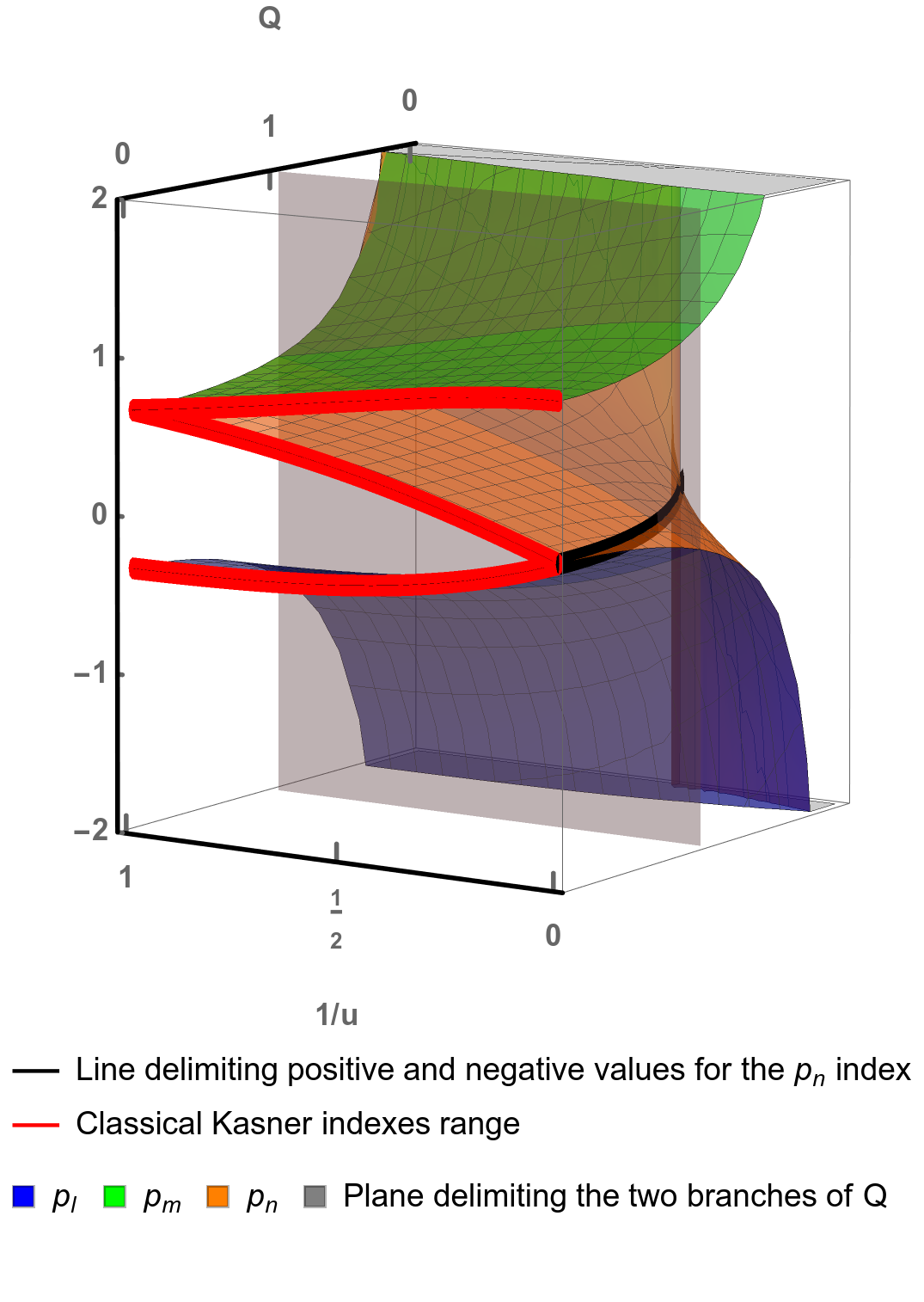}
        \end{center}
        \caption{Plot of the ``spectrum'' of the Kasner indices in 
                the $(u,Q)$-parametrization~\eqref{parametrization:u-Q-param}. Every
                value of $(u,Q)$ in the dominion $u\in\lbrack 1,\infty)$ and
                $Q\in[0,1]$ uniquely selects a triplet of Kasner indices. The
                vertical plane delimits the two branches of the
                $(u,Q)$-parametrization~\eqref{parametrization:u-Q-param}. The
                connected-branch is the one containing the red (dark gray) lines, i.e.\ 
                the
                standard parametrization shown in Fig.~\ref{fig:kasner:u-param}.
                The $Q$ parameter of the connected-branch is increasing from 0 to
                1 in the positive direction of the $Q$-axis, while for the
                disconnected-branch it is decreasing from 1 to 0.  The black line
                delimits the part of the $(u,Q)$-plane where there is only one
                (above) or two (below) negative Kasner indices. The plot is cut on
                the bottom and on top but it is actually extending up to
                $\pm\infty$.}
        \label{fig:parametrization:u-Q}
\end{figure}

The $(u,Q)$-parametrization~\eqref{parametrization:u-Q-param} is even
in $Q$: $p_a(u,Q) = p_a(u,-Q)$ with $a=l,m,n$. We can thus assume
without loss of generality that $0 \leq Q \leq 1$. Another interesting
feature of the Polymer Bianchi I model, that is evident from
Fig.~\ref{fig:parametrization:u-Q}, is that, for every $u$ and $Q$ in
a non-null measure set, two Kasner indices can be simultaneously
negative (instead of only one as is the case of the standard
$u$-parametrization~\eqref{kasner:u-param}).

In a similar fashion as the $u$ parameter is ``remapped'' in the
standard case (second line of~\eqref{bianchi-ix:BKL-map}), also the
$(u,Q)$ parametrization is to be ``remapped'', if the $u$ parameter
happens to become less than one. Down below we list these remapping
prescriptions explicitly: we show how the
$(u,Q)$-parametrization~\eqref{parametrization:u-Q-param} can be
recovered if the $u$ parameter becomes smaller than 1, through a
reordering of the Kasner indices and a remapping of the $u$ parameter.
\begin{itemize}\label{parametrization:remapping}
        \item \centering $0 < u < 1$
        \[\begin{dcases} p_1\left(u,Q\right) \rightarrow
        p_1\left(\frac{1}{u},Q\right)\\ p_2\left(u,Q\right) \rightarrow
        p_3\left(\frac{1}{u},Q\right)\\ p_3\left(u,Q\right) \rightarrow
        p_2\left(\frac{1}{u},Q\right)
        \end{dcases}\]
        \item \centering $-\tfrac{1}{2} < u < 0$
        \[\begin{dcases} p_1\left(u,Q\right) \rightarrow
        p_2\left(-\frac{1+u}{u},Q\right)\\ p_2\left(u,Q\right)
        \rightarrow p_3\left(-\frac{1+u}{u},Q\right)\\
        p_3\left(u,Q\right) \rightarrow p_1\left(-\frac{1+u}{u},Q\right)
        \end{dcases}\]
        \item \centering $-1 < u < -\frac{1}{2}$
        \[\begin{dcases} p_1\left(u,Q\right) \rightarrow p_3\left(
        -\frac{u}{1+u},Q\right)\\ p_2\left(u,Q\right) \rightarrow
        p_2\left( -\frac{u}{1+u},Q\right)\\ p_3\left(u,Q\right)
        \rightarrow p_1\left( -\frac{u}{1+u},Q\right)
        \end{dcases}\]
        \item \centering $-2 < u < -1$
        \[\begin{dcases} p_1\left(u,Q\right) \rightarrow p_3\left(
        -\frac{1}{1+u},Q\right)\\ p_2\left(u,Q\right) \rightarrow
        p_1\left( -\frac{1}{1+u},Q\right)\\ p_3\left(u,Q\right)
        \rightarrow p_2\left( -\frac{1}{1+u},Q\right)
        \end{dcases}\]
        \item \centering $u < -2$
        \[\begin{dcases} p_1\left(u,Q\right) \rightarrow p_2\left(
        -\left(1+u\right),Q\right)\\ p_2\left(u,Q\right) \rightarrow
        p_1\left( -\left(1+u\right),Q\right)\\ p_3\left(u,Q\right)
        \rightarrow p_3\left( -\left(1+u\right),Q\right)
        \end{dcases}\]
\end{itemize}
where the indices on the left are defined for the $u$ shown after the
bullet $\bullet$, while the indices on the right are in the
``correct'' range $u>1$. The arrow `$\rightarrow$' is there to
indicate a conceptual mapping. However, in value it reads as an
equality `$=$'. As far as the Q parameter is concerned, a
``remapping'' prescription for $Q$ is not needed because values
outside the boundaries~\eqref{polymer-bianchi-i:Q-range} are not
physical and should never be considered.

In Figure~\ref{fig:parametrization:u-Q} the values of the ordered
Kasner indices are displayed for the
$(u,Q)$-parametrization~\eqref{parametrization:u-Q-param}, where
$Q\in\left\lbrack 0,1 \right\rbrack$. Because the range $u > 1$ is not
easily plottable, the equivalent parametrization in
$u\in\left\lbrack 0,1 \right\rbrack$ was used. We notice that the
roles of $p_m$ and $p_n$ are exchanged for this range choice.

\subsection{Polymer Bianchi II}\label{sec:polymer-bianchi-ii}

Here we apply the method described
in~\ref{sec:polymer-bianchi-i-dynamics} to find an approximate
solution to the Einstein's Eqs of the Polymer Bianchi II model. We
start by selecting the $V_B$ potential appropriate for Bianchi
II~\eqref{chiara:BianchiIIpotential} and we substitute it in the
Hamiltonian~\eqref{chiara:polymerSuperHam} (in the following we will
always assume the time gauge $N=1$).

Then, starting from Hamilton's Eqs, inverting the Misner canonical
transformation and converting the synchronous time $t$-derivatives
into logarithmic time $\tau$-derivatives, the Polymer Bianchi II
Einstein's Eqs are found to be:
\begin{equation}\label{polymer-bianchi-ii:einstein-equations}
\begin{dcases}
{q_l}_{\tau\tau}(\tau) = \frac{1}{3} e^{2q_l(\tau)} \left\lbrack
-4 + \sqrt{1+{\left( \frac{2{(4\pi)}^2\mu}{\kappa} v(\tau)
                \right)}^2 } \right\rbrack \\
{q_m}_{\tau\tau}(\tau) = \frac{1}{3} e^{2q_l(\tau)} \left\lbrack 2
+ \sqrt{1+{\left( \frac{2{(4\pi)}^2\mu}{\kappa} v(\tau)
                \right)}^2 } \right\rbrack \\
{q_n}_{\tau\tau}(\tau) = \frac{1}{3} e^{2q_l(\tau)} \left\lbrack 2
+ \sqrt{1+{\left( \frac{2{(4\pi)}^2\mu}{\kappa} v(\tau)
                \right)}^2 } \right\rbrack
\end{dcases}
\end{equation}
where
$v(\tau) := {q_l}_\tau(\tau) + {q_m}_\tau(\tau) +
{q_n}_\tau(\tau)$. In the $Q \ll 1$ approximation it is enough to
consider only the connected-branch of the solutions, i.e.~the branch
that in the limit $\mu\to 0$ make
Eqs~\eqref{polymer-bianchi-ii:einstein-equations} reduce to the
correct classical Eqs~\eqref{bianchi-ii:log-einstein}.

Now we find an approximate solution to the
system~\eqref{polymer-bianchi-ii:einstein-equations}, with the only
assumption that $\mu$ is little compared to the cubic root of the
Universe volume.  We are entitled then to exploit the standard
perturbation theory and expand the solution until the first non-zero
order in $\mu$. Because $\mu$ appears only squared $\mu^2$ in the
Einstein's Eqs~\eqref{polymer-bianchi-ii:einstein-equations}, the
perturbative expansion will only contain even powers of $\mu$.

First we expand the solution at the first order in $\mu^2$
\begin{equation}\label{polymer-bianchi-ii:approximate-solution}
\begin{dcases}
q_l(\tau) = q_l^0(\tau) + \mu^2 q_l^1(\tau) + O(\mu^4) \\
q_m(\tau) = q_m^0(\tau) + \mu^2 q_m^1(\tau) + O(\mu^4) \\
q_n(\tau) = q_n^0(\tau) + \mu^2 q_n^1(\tau) + O(\mu^4)
\end{dcases}
\end{equation}
where $q_{l,m,n}^0$ are the zeroth order terms and $q_{l,m,n}^1$ are
the first order terms in $\mu^2$.  We recall the zeroth order solution
$q_{l,m,n}^0$ is just the classical
solution~\eqref{bianchi-ii:solution}, where the $q_{l,m,n}(\tau)$
there must be now appended a 0-apex, accordingly.

Considering the fact that ${q_m}_{\tau\tau} = {q_n}_{\tau\tau}$, we
can simplify $q_n(\tau)$
from~\eqref{polymer-bianchi-ii:einstein-equations}, by setting
\begin{multline}\label{approximate-solution:perturbative-expansion}
q_n(\tau) = q_m(\tau) + 2 (c_5 + c_6\tau) + \mu^2 (c_{11} + c_{12}
\tau) + \\ - 2 (c_3 + c_4\tau) - \mu^2 (c_9 + c_{10}\tau)
\end{multline}
where the first six constants of integration $c_1,c_2,\dots,c_{6}$
play the same role at the zeroth order as in
Eq.~\eqref{bianchi-ii:solution}, while the successive six
$c_7,\dots,c_{12}$ are needed to parametrize the first order
solution.

Then we substitute~\eqref{approximate-solution:perturbative-expansion}
in~\eqref{polymer-bianchi-ii:einstein-equations} and, as it is
required by perturbations theory, we gather only the zeroth and first
order terms in $\mu^2$ and neglect all the higher order terms. The
Einstein's Eqs for the first order terms $q_l^1$ and $q_m^1$ are then
found to be:
\begin{numcases}{\label{approximate-solution:1order-eqs-2}}
\label{approximate-solution:1order-eqs-2-ql}
\frac{\partial^2{q_l^1}}{\partial\tau^2} + {c_1}^2 \sech^2(c_1\tau +
c_2) \times \\ \nonumber \times \left\{ 2 {q_l^1} + \frac{C}{3}
{\left\lbrack 2(c_6+c_4) + c_1 \tanh(c_1\tau + c_2)
        \right\rbrack}^2 \right\} = 0\\
\label{approximate-solution:1order-eqs-2-qm}
\frac{\partial^2{q_m^1}}{\partial\tau^2} + {c_1}^2 \sech^2(c_1\tau +
c_2) \times \\ \nonumber \times \left\{ 2 {q_l^1} + \frac{C}{3}
{\left\lbrack 2(c_6+c_4) + c_1 \tanh(c_1\tau + c_2)
        \right\rbrack}^2 \right\} = 0
\end{numcases}
where we have defined $C \equiv \frac{2{(4\pi)}^2}{\kappa^2}$ for
brevity and we have substituted the zeroth order
solution~\eqref{bianchi-ii:solution} where needed.

Eqs~\eqref{approximate-solution:1order-eqs-2} are two almost uncoupled
non-homogeneous ODEs.\@ \eqref{approximate-solution:1order-eqs-2-ql}
is a linear second order non-homogeneous ODE with non-constant
coefficients and, being completely uncoupled, it can be solved
straightly.\@ \eqref{approximate-solution:1order-eqs-2-qm} is solved
by mere substitution of the solution
of~\eqref{approximate-solution:1order-eqs-2-ql} in it and subsequent
double integration.

To solve~\eqref{approximate-solution:1order-eqs-2-ql} we exploited a
standard method that can be found, for example, in~\cite[Lesson
23]{Tenenbaum:2012}. This method is called \emph{reduction of order
        method} and, in the case of a second order ODE, can be used to find
a particular solution once any non trivial solution of the related
homogeneous equation is known.

The homogeneous equation associated
to~\eqref{approximate-solution:1order-eqs-2-ql} is
\begin{equation}\label{approximate-solution:homogeneous-eq}
\frac{\partial^2{q_l^1}}{\partial\tau^2} + 2{c_1}^2 \sech^2(c_1\tau
+ c_2) {q_l^1} = 0
\end{equation}
whose general solution reads as:
\begin{equation}\label{approximate-solution:homogeneous-solution}
{q_l^1}_O(\tau) = -\frac{c_8}{c_1} + (c_7 + c_8 \tau) \tanh(c_1\tau
+ c_2)
\end{equation}

By applying the above-mentioned method, we obtain the following
solution for~\eqref{approximate-solution:1order-eqs-2-ql}
\begingroup\makeatletter\def\f@size{9}\check@mathfonts
\def\maketag@@@#1{\hbox{\m@th\large\normalfont#1}}
\begin{equation}\label{approximate-solution:1order-q-l-sol}
\begin{split}
q_l^1(\tau) = & -\tfrac{c_8}{c_1} + (c_7 + c_8 \tau) \tanh(c_1\tau
+
c_2)\\
& -\tfrac{C}{36}c_1 \tanh(c_1\tau + c_2) \left\{ 3 \left\lbrack
{c_1}^2 + 8 {(c_4 + c_6)}^2 \right\rbrack \tau \right. \\
& + 16 (c_4 + c_6) \ln\left\lbrack\cosh(c_1\tau +
c_2)\right\rbrack - 3c_1 \tanh(c_1\tau + c_2) \Big\}
\end{split}
\end{equation}
\endgroup The solution
for~\eqref{approximate-solution:1order-eqs-2-qm} is found by
substituting~\eqref{approximate-solution:1order-q-l-sol} in it and
then integrating two times:
\begingroup\makeatletter\def\f@size{8.8}\check@mathfonts
\def\maketag@@@#1{\hbox{\m@th\large\normalfont#1}}
\begin{equation}
\begin{split}
q_m^1(\tau) = &
~c_{10} + c_9\tau - (c_7 + c_8 \tau) \tanh(c_1 \tau + c_2)\\
& - \tfrac{C}{36} \Big\{ 16 c_1 (c_4 + c_6)(c_1\tau + c_2)
{c_1}^2\sech^2(c_1\tau + c_2)  \\
& + 8 \left\lbrack {c_1}^2 + 12 {(c_4+c_6)}^2 \right\rbrack \ln
(\cosh (c_1 \tau +
c_2)) \\
& - c_1\tanh(c_1 \tau + c_2) \left\lbrack 48 (c_4 + c_6) + 3
\left({c_1}^2 + 8{(c_4 + c_6)}^2\right)\tau
\right. \\
& + 16(c_4 + c_6) \ln(\cosh(c_1\tau + c_2)) \Big\rbrack \Big\}
\end{split}
\end{equation}
\endgroup
Now that we know $q_l^0$, $q_l^1$, $q_m^0$ and $q_m^1$, the
complete solution for $q_l$, $q_m$ and $q_n$ is found
through~\eqref{polymer-bianchi-ii:approximate-solution}
and~\eqref{approximate-solution:perturbative-expansion} by mere
substitution.

\section{Polymer BKL map}\label{sec:polymer-BKL-map}
In this last Section, we calculate the Polymer modified BKL map
and study some of its properties.

In Sec.~\ref{sec:polymer-bkl-map-kasner-indices} the Polymer BKL map
on the Kasner indices is derived while in
Sec.~\ref{sec:BKL-map-properties} some noteworthy properties of the
map are discussed.

Finally in Sec~\ref{sec:numerical-simulation} the results of a simple
numerical simulation of the Polymer BKL map over many iterations are
presented and discussed.

\subsection{Polymer BKL map on the Kasner indices}
\label{sec:polymer-bkl-map-kasner-indices}

Here we use the method outlined in
Secs. from~\ref{sec:polymer-bianchi-i-dynamics} to
\ref{sec:polymer-bianchi-ii} to directly calculate the Polymer BKL map
on the Kasner indices.

First, we look at the asymptotic limit at $\pm\infty$ for the solution
of Polymer Bianchi
II~\eqref{polymer-bianchi-ii:approximate-solution}. As in the standard
case, the Polymer Bianchi II model ``links together'' two Kasner
epochs at plus and minus infinity. In this sense, the dynamics of
Polymer Bianchi II is not qualitatively different from the standard
one. We will tell the quantities at plus and minus infinity apart by
adding a prime $\Box'$ to the quantities at minus infinity and leaving
the quantities at plus infinity un-primed. The two Kasner solutions at
plus and minus infinity can be still parametrized according
to~\eqref{bianchi-i:solution}.

By summing together equations~\eqref{bianchi-i:solution} and using the
first Kasner constraint~\eqref{kasner:sum}, we find that:
\begin{equation}\label{polymer-BKL-map:lambda}
\begin{cases}
\lim_{\tau \rightarrow +\infty} \frac{1}{2\tau}\left( q_l + q_m +
q_n \right) =
\Lambda \\
\lim_{\tau \rightarrow -\infty} \frac{1}{2\tau}\left( q'_l + q'_m
+ q'_n \right) = \Lambda'
\end{cases}
\end{equation}

By taking the limits at $\tau \rightarrow \pm \infty$ of the
derivatives of the Polymer Bianchi II
solution~\eqref{polymer-bianchi-ii:approximate-solution},\\~\\
{\centering $\boxed{\lim_{\tau \rightarrow +\infty}}$\par}
\begingroup\makeatletter\def\f@size{8.8}\check@mathfonts
\def\maketag@@@#1{\hbox{\m@th\large\normalfont#1}}
\begin{align}
  \nonumber {q_l}_\tau = & - c_1 + \mu^2 c_8 - \mu^2 \frac{C}{36} c_1
                           \left\lbrack 3{c_1}^2 + 16 c_1 (c_4 + c_6)
                           + 24 {(c_4 + c_6)}^2 \right\rbrack \\
  \nonumber {q_m}_\tau = & ~ 2c_4 + c_1 + \mu^2 (-c_8 + c_9) - \mu^2
                           \frac{C}{36} c_1 \left\lbrack 5{c_1}^2 + 24 {(c_4 +
                           c_6)}^2 \right\rbrack \\
  \nonumber {q_n}_\tau = & ~ 2c_6 + c_1 + \mu^2 (-c_8 + c_{11}) - \mu^2
                           \frac{C}{36} c_1 \left\lbrack 5{c_1}^2 + 24 {(c_4 +
                           c_6)}^2 \right\rbrack \\
  \nonumber \Lambda = & ~ \frac{c_1}{2} + c_4 + c_6 + \mu^2 \half (-c_8 + c_9 +
                        c_{11}) + \\ 
                         & ~ - \mu^2 \frac{C}{72} c_1 \left\lbrack
                           13 {c_1}^2 + 16 c_1 (c_4 + c_6) + 168 {(c_4 + c_6)}^2
                           \right\rbrack \label{polymer-BKL-map:limit-plus}
\end{align}
\endgroup
{\centering$\boxed{\lim_{\tau \rightarrow -\infty}}$\par}
\begingroup\makeatletter\def\f@size{8.8}\check@mathfonts
\def\maketag@@@#1{\hbox{\m@th\large\normalfont#1}}
\begin{align}
  \nonumber {q'_l}_\tau = & ~ c_1 - \mu^2 c_8 + \mu^2 \frac{C}{36} c_1
                            \left\lbrack 3{c_1}^2 - 16 c_1 (c_4 + c_6)
                            + 24 {(c_4 + c_6)}^2 \right\rbrack \\
  \nonumber {q'_m}_\tau = & ~ 2c_4 - c_1 - \mu^2 (c_8 + c_9) + \mu^2
                            \frac{C}{36} c_1 \left\lbrack 5{c_1}^2 + 24 {(c_4 +
                            c_6)}^2 \right\rbrack \\
  \nonumber {q'_n}_\tau = & ~ 2c_6 - c_1 - \mu^2 (c_8 + c_{11}) + \mu^2
                            \frac{C}{36} c_1 \left\lbrack 5{c_1}^2 + 24 {(c_4 +
                            c_6)}^2 \right\rbrack \\
  \nonumber \Lambda' = & ~ -\frac{c_1}{2} + c_4 + c_6 + \mu^2 \half (c_8 + c_9 +
                         c_{11}) + \\ 
                          & ~ + \mu^2 \frac{C}{72} c_1 \left\lbrack
                            13 {c_1}^2 - 16 c_1 (c_4 + c_6) + 168 {(c_4 + c_6)}^2
                            \right\rbrack \label{polymer-BKL-map:limit-minus}
\end{align}
\endgroup
and comparing~\eqref{polymer-BKL-map:limit-plus}
and~\eqref{polymer-BKL-map:limit-minus}
with~\eqref{bianchi-i:solution} and~\eqref{polymer-BKL-map:lambda}, we
can find the following expressions for the primed and unprimed Kasner
indices and $\Lambda$:
\begin{numcases}{\label{polymer-BKL-map:system}}
2 \Lambda{p_l}  = f_l (\boldsymbol{c})\\
2 \Lambda{p_m}  = f_m (\boldsymbol{c})\\
2 \Lambda{p_n}  = f_n (\boldsymbol{c})\\
\Lambda= f_\Lambda(\boldsymbol{c})\\
\label{polymer-BKL-map:rel1Minus}
2 \Lambda'{p'_l} = f'_l (\boldsymbol{c})\\
\label{polymer-BKL-map:rel2Minus}
2 \Lambda'{p'_m} = f'_m (\boldsymbol{c})\\
2 \Lambda'{p'_n} = f'_n (\boldsymbol{c})\\
\label{polymer-BKL-map:rel4Minus}
\Lambda' = f'_\Lambda(\boldsymbol{c})
\end{numcases}
where the functions $f_{l,m,n,\Lambda}$ and $f'_{l,m,n,\Lambda}$
correspond to the r.h.s.\ of~\eqref{polymer-BKL-map:limit-plus}
and~\eqref{polymer-BKL-map:limit-minus} respectively and
$\boldsymbol{c}$ is a shorthand for the set $\{c_1,\dots,c_{12}\}$.

Now, in complete analogy with~\eqref{bianchi-ii:asympt-rel}, we look
for an asymptotic condition. We don't need to solve the whole
system~\eqref{polymer-BKL-map:system}, but only to find a relation
between the old and new Kasner indices and $\Lambda$ at the first
order in $\mu^2$. In practice, not all of
the~\eqref{polymer-BKL-map:system} relations are actually needed to
find an asymptotic condition.

We recall from~\eqref{bianchi-ii:BKL-map} that the standard BKL map on
$\Lambda$ is
\begin{equation}\label{polymer-BKL-map:classical-map-on-lambda}
\Lambda' = (1 + 2 p_l) \Lambda
\end{equation}
where we have assumed that $p_l < 0$, as we will continue to do in the
following. As everything until now is hinting to, we prescribe that
the Polymer modified BKL map reduces to the standard BKL map when
$\mu\to 0$, so that
relation~\eqref{polymer-BKL-map:classical-map-on-lambda} is modified
in the Polymer case only perturbatively. Since we are considering
only the first order in $\mu^2$, we can write for the Polymer case:
\begin{equation}
\Lambda' = (1 + 2 p_l) \Lambda + \mu^2 h(\boldsymbol{c}) + O(\mu^4)
\end{equation}
Solving for $h(\boldsymbol{c})$, we find that
\begin{align*}
h(\boldsymbol{c}) & = \frac{1}{\mu^2} \left( \Lambda' - \Lambda - 2p_l
\Lambda \right)\\ 
& = \frac{1}{\mu^2} \left( f'_\Lambda(\boldsymbol{c})
- f_\Lambda(\boldsymbol{c}) - f_l(\boldsymbol{c}) \right)\\
& = \frac{4C}{9} c_1 \left\lbrack {c_1}^2 + c_1 (c_4 + c_6) + 12
{(c_4 + c_6)}^2 \right\rbrack
\end{align*}
We can invert the subsystem made up by the zeroth order of
equations~\eqref{polymer-BKL-map:rel1Minus},
~\eqref{polymer-BKL-map:rel2Minus}
and~\eqref{polymer-BKL-map:rel4Minus}
\begin{equation*}
\begin{dcases}
2 \Lambda{p_l} = -c_1\\
2 \Lambda{p_m} = 2c_4 + c_1\\
\Lambda = \frac{c_1}{2} + c_4 + c_6
\end{dcases}
\end{equation*}
to express $h=h(p_l,p_m,\Lambda)$. Finally the asymptotic condition
reads as:
\begin{equation}\label{polymer-BKL-map:asymptotic-condition}
\Lambda' \approx (1 + 2 p_l) \Lambda - \mu^2 \frac{16C}{9}  
{\Lambda}^3 p_l \left( 6 + 11p_l + 7{p_l}^2 \right)
\end{equation}
We stress that this is only one out of many equivalent ways to extract
from the system~\eqref{polymer-BKL-map:system} an asymptotic
condition.

Now, we need other three conditions to derive the Polymer BKL map. One
is provided by the sum of the primed Kasner indices at minus infinity.
This is the very same both in the standard case~\eqref{kasner:sum} and
in the Polymer case.

Sadly, the two conditions~\eqref{bianchi-ii:rel-1}
and~\eqref{bianchi-ii:rel-2} are not valid anymore. Instead, one
condition can be derived by noticing that
\begin{gather*}
{q_m}_{\tau\tau} - {q_n}_{\tau\tau} = 0 ~~ \Rightarrow ~~
{q_m}_{\tau} - {q_n}_{\tau} = \text{const} ~~ \Rightarrow ~~\\
({q_m}_{\tau} - {q_n}_{\tau})|_{\tau\rightarrow +\infty} =
({q_m}_{\tau} - {q_n}_{\tau})|_{\tau \rightarrow -\infty} ~~ \Rightarrow ~~\\
\Lambda(p_l - p_m) = \Lambda'(p'_l - p'_m)
\end{gather*}

Lastly, we choose as the fourth condition the sum of the squares of
the Kasner indices at minus
infinity~\eqref{polymer-bianchi-i:sumofsquares-exact}. Because of the
assumption $Q \ll 1$, we will consider here only the connected branch
of~\eqref{polymer-bianchi-i:sumofsquares-exact}. We gather now all the
four conditions and put them in a system:
\begin{equation}\label{polymer-BKL-map:4-conditions}
\begin{dcases}
p'_l + p'_m + p'_n = 1\\
Q (p_l - p_m) = Q' (p'_l - p'_m)\\
Q' \approx (1 + 2 p_l) Q - \tfrac{2}{9} Q^3 p_l \left(
6 + 11p_l + 7{p_l}^2 \right) \\
{p'_l}^2 + {p'_m}^2 + {p'_n}^2 = \frac{1}{3} \left( 1 + \frac{4}{1
        + \sqrt{1 - Q^2 }} \right)
\end{dcases}
\end{equation}
where we have also used definition~\eqref{polymer-bianchi-i:Q}.

Finding the polymer BKL map is now only a matter of solving the
system~\eqref{polymer-BKL-map:4-conditions} for the un-primed
indices. The \emph{Polymer BKL map} at the first order in $\mu^2$ is
then:
\begin{equation}\label{polymer-BKL-map:polymer-BKL-map-kasner}
\begin{split}
p'_l & \approx - \frac{p_l}{1 + 2p_l} - \frac{2}{9} Q^2
\left\lbrack \frac{7 + 14p_l + 9{p_l}^2}{{(1 + 2p_l)}^2}
\right\rbrack\\
p'_m & \approx \frac{2p_l + p_m}{1 + 2p_l} + \frac{2}{9} Q^2 p_l \times \\
& \times \left\lbrack \frac{ -3 + p_l + 9{p_l}^2 + 8{p_l}^3 + p_m
        \left( 6 + 11p_l + 7{p_l}^2 \right) }{{(1 + 2p_l)}^2}
\right\rbrack\\
p'_n & \approx \frac{2p_l + p_n}{1 + 2p_l} + \frac{2}{9} Q^2 p_l \times \\
& \times \left\lbrack \frac{ -3 + p_l + 9{p_l}^2 + 8{p_l}^3 + p_n
        \left( 6 + 11p_l + 7{p_l}^2 \right) }{{(1 + 2p_l)}^2}
\right\rbrack\\
Q' & \approx (1 + 2 p_l) Q - \tfrac{2}{9} Q^3 p_l \left(
6 + 11p_l + 7{p_l}^2 \right)\\
\end{split}
\end{equation}
where all the terms of order $O(Q^4)$ were neglected.  It is worth
noticing that, by taking the limit $\mu \rightarrow 0$, the standard
BKL map~\eqref{bianchi-ii:BKL-map} is immediately recovered. We stress
that the form of the map is not unique.  One can use the two Kasner
constraints~\eqref{kasner:sum}
and~\eqref{polymer-bianchi-i:sumofsquares-exact} to ``rearrange'' the
Kasner indices as needed.

\subsection{BKL map properties}\label{sec:BKL-map-properties}

First we derive how the Polymer BKL map can be expressed in terms of
the $u$ and $Q$ parameters:
\begin{equation}\label{BKL-map-properties:BKL-map-sketch}
  \begin{dcases}
    u\\
    Q
  \end{dcases}
  \xrightarrow[\text{map}]{\text{BKL}}
  \begin{dcases}
    u' = u'(u,Q)\\
    Q' = Q'(u,Q)\\
  \end{dcases}
\end{equation}
Then we discuss some of its noteworthy properties.

We start by inserting in the polymer BKL
map~\eqref{polymer-BKL-map:polymer-BKL-map-kasner} the connected
branch of the $(u,Q)$
parametrization~\eqref{parametrization:u-Q-param} and requiring the
relations
\begin{equation}
  \begin{dcases}
    p_l = p_1(u,Q)\\
    p_m = p_2(u,Q)\\
    p_n = p_3(u,Q)
  \end{dcases}
  \xrightarrow[\text{map}]{\text{BKL}}
  \begin{dcases}
    p'_l = p_2(u'(u,Q),Q'(u,Q))\\
    p'_m = p_1(u'(u,Q),Q'(u,Q))\\
    p'_n = p_3(u'(u,Q),Q'(u,Q))
  \end{dcases}
\end{equation}
to be always satisfied. The resulting Polymer BKL map on $(u,Q)$ is
quite complex. We therefore split it in many terms:
\begin{align*}
  A(u) = & ~ 72 u {\left(u^2 + u + 1 \right)}^2 \left( -2 +3u -3u^2 +u^3
           \right)\\
  B(u) = & ~ -12 - 6 u + 53 u^2 - 119 u^3 + 204 u^4 + \\ 
         & ~ - 187 u^5 + 112 u^6 - 57 u^7 + 3 u^8 \\
  C(u) = & ~ 5184 {(1 + u^2 + u^4)}^4\\
  D(u) = & ~ 1296 {(1 - u + u^2)}^6 {(1 + u + u^2)}^2\\
  E(u) = & ~ 3 {(1 - u + u^2)}^2 \left(63 - 162 u + 663 u^2 - 1350 u^3 + \right. \\ 
         & ~ + 2398 u^4 - 3402 u^5 + 3607 u^6 - 3402 u^7 + \\
         & \left. + 2398 u^8 - 1350 u^9 + 663 u^{10} - 162 u^{11} + 63 u^{12} \right)\\
  F(u) = & ~ 72 (-1 + u + 3 u^3 + 3 u^5 + u^6 + 2 u^7)\\
  G(u) = & ~ -3 + 57 u - 112 u^2 + 187 u^3 - 204 u^4 + \\
         & ~ + 119 u^5 - 53 u^6 + 6 u^7 + 12 u^8 \\
  H(u) = & ~ -72 {(1 + u^2 + u^4)}^2 \\
  L(u) = & ~ 3 \left( 1 + u^2 + u^4 \right)
\end{align*}
that have to be inserted in
\begin{subequations}\label{BKL-map-properties:BKL-map}
  \begin{align}
    \label{BKL-map-properties:BKL-map-u}
    u' & = \frac{ A(u) + Q^2 B(u) + \sqrt{C(u) + Q^2 D(u) + Q^4
         E(u)}}{F(u) + Q^2 G(u)} \\
    \label{BKL-map-properties:BKL-map-Q}
    Q' & = \frac{\sqrt{ H(u) + \sqrt{C(u) + Q^2 D(u) + Q^4 E(u)
         }}}{L(u)}
  \end{align}
\end{subequations}
where we recall that the dominions of definition for $u$ and $Q$ are
$u\in\lbrack 1,\infty)$ and $Q\in[0,1]$.  All the square-roots and
denominators appearing in~\eqref{BKL-map-properties:BKL-map} are well
behaved (always with positive argument or non zero respectively) for
any $(u,Q)$ in the intervals of definition. It is not clearly evident
at first sight, but the polymer BKL map reduces to the standard
one~\eqref{bianchi-ix:BKL-map} if $Q \rightarrow 0$.
                
Now, we study the asymptotic behavior of the polymer BKL
map~\eqref{BKL-map-properties:BKL-map-u} for $u$ going to
infinity. This limit is relevant to us for two reasons. Firstly, In
Sec.~\ref{sec:semiclassical-analysis} we already pointed out that
infinitely many bounces of the $\beta$-point against the potential
walls happen until the singularity is reached. This means that the
Polymer BKL map is to be iterated infinitely many times.
                
Secondly, we suppose the Polymer BKL map (or at least its $u$
portion~\eqref{BKL-map-properties:BKL-map-u}) to be ergotic, so that
every open set of the parameter space is visited with non null
probability. This assumption is backed up by the observation that the
Polymer BKL map~\eqref{BKL-map-properties:BKL-map-u} tends
asymptotically to the standard BKL map~\eqref{bianchi-ix:BKL-map}
(that is ergodic), as shown in the following
Sec.~\ref{sec:numerical-simulation} through a numerical simulation.
                
\begin{figure}
  \begin{center}
    \includegraphics[width=0.85\columnwidth]{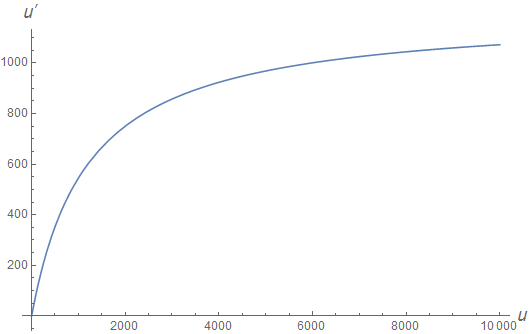}
  \end{center}
  \caption{The Polymer BKL map for
    $u$~\eqref{BKL-map-properties:BKL-map-u} is plotted for
    $Q = 1/10$.  There a plateau $u' \approx 1200$ is reached at about
    $u \approx 10^4$. This means that for any $u \gtrsim 10^4$ the
    next $u'$ will inevitably be $u' \approx 1200$, no matter how big
    $u$ is.}
  \label{BKL-map-properties:u-map}
\end{figure}

Hence, every open interval of the $u > 1$ line is to be visited
eventually: we want therefore to assess how the map behaves for
$u\to\infty$. Physically, a big value for $u$ means a long (in term of
epochs) Kasner era, and in turn this means that the Universe is going
deep inside one of the corners of Fig.~\ref{fig:potential}. As we can
appreciate from Fig.~\ref{BKL-map-properties:u-map}, for
$u\rightarrow\infty$, $u'$ reaches a plateau:
\begin{equation}\label{BKL-map-properties:plateau}
  \lim_{u\rightarrow\infty} u'(u,Q) = \frac{24 + Q^2 + \sqrt{3\left(7
        Q^4+48 Q^2+192\right)}}{4 Q^2}
\end{equation}
The plateau~\eqref{BKL-map-properties:plateau} is higher and steeper
the more $Q$ is close to zero. This physically means that there is a
sort of ``centripetal potential'' that is driving the $\beta$-point
off the corners and towards the center of the triangle.  We infer that
this ``centripetal potential'' is somehow linked to the velocity-like
quantity $v(\tau)$ that appears in the polymer Bianchi II Einstein's
Eqs~\eqref{polymer-bianchi-ii:einstein-equations}. Because of this
plateau, the mechanism that drives the Universe away from the corner,
implicit in the standard BKL map~\eqref{bianchi-ix:BKL-map}, seems to
be much more efficient in the quantum case.  For $Q\rightarrow 0$, the
plateau tends to disappear:
$\lim_{\substack{u\rightarrow\infty\\Q\rightarrow 0}} u'(u,Q) =
\infty$.
                
As a side note, we remember that it has not been proved analytically
that the standard BKL map is still valid deep inside the corners. As a
matter of fact, the BKL map can be derived analytically only for the
very center of the edges of the triangle of Fig.~\ref{fig:potential},
because those are the only points where the Bianchi IX potential is
exactly equal to the Bianchi II potential. The farther we depart from
the center of the edges, the more the map looses precision. At any
rate, there are some numerical studies for the Bianchi IX
model~\cite{Moser:1973,Berger:2002} that show how the standard BKL map
is valid with good approximation even inside the corners.
                
The analysis of the behavior of the Polymer BKL map for the $Q$
parameter~\eqref{BKL-map-properties:BKL-map-Q} is more
convoluted. Little can be said analytically about the overall behavior
across multiple iterations because of its evident complexity. For this
reason, in Sec.~\ref{sec:numerical-simulation} we discuss the results
of a simple numerical simulation that probes the behavior of the
map~\eqref{BKL-map-properties:BKL-map} over many iterations.
                
The most important point to check is if the dominion of
definition~\eqref{polymer-bianchi-i:Q-range} for $Q$ is preserved by
the map. We remember that for $Q \approx 1$ the perturbation theory,
by the means of which we have derived the map, is not valid
anymore. So every result in that range is to be taken with a grain of
salt.
                
\begin{figure}
  \begin{center}
    \includegraphics[width=0.7\columnwidth]{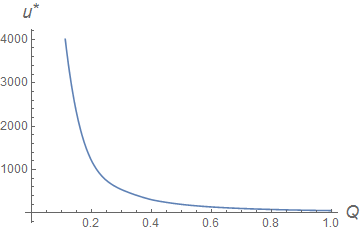}
  \end{center}
  \caption{For any $Q\in[0,1]$ it is shown the
    corresponding $u^*(Q)$ such that for $u < u^*$ the Polymer BKL map
    on $Q$~\eqref{BKL-map-properties:BKL-map-Q} is monotonically
    decreasing. For a single iteraction it can be shown that for every
    $Q < 0.96$ the following $Q' < 1$. The function $u^*(Q)$ cannot be
    given in closed form.}
  \label{BKL-map-properties:threshold}
\end{figure}

In Fig.~\ref{BKL-map-properties:threshold} it is shown the maximum
value of $u^*=u^*(Q)$ for which the polymer BKL map on $Q$ is
monotonically decreasing, i.e.\ for any $Q\in[0,1]$ it is shown the
corresponding $u^*$, such that for $u < u^*$ the polymer BKL map on
$Q$ is monotonically decreasing. It is clear how the more $Q$ is
little the more $u$ can grow before coming to the point that $Q$ stops
decreasing and starts increasing.
                
That said, as soon as we consider the behavior of $u$, and
particularly the plateau of Fig.~\ref{BKL-map-properties:u-map}, we
notice that the Universe cannot ``indulge'' much time in ``big-$u$
regions'': even if it happens to assume a huge value for $u$, this is
immediately dampened to a much smaller value at the next
iteration. The net result is that $Q$ is almost always decreasing as
it is also strongly suggested by the numerical simulation of
Sec.~\ref{sec:numerical-simulation}.
                
Summarizing, the Polymer BKL map on
$Q$~\eqref{BKL-map-properties:BKL-map-Q}, apart from a small set of
initial conditions in the region where the perturbation theory is
failing $Q \approx 1$, preserves the dominion of definition of
$Q$~\eqref{polymer-bianchi-i:Q-range} and is decreasing at almost any
iteration.
                
\subsection{Numerical simulation}\label{sec:numerical-simulation}

In this Section we present the results of a simple numerical
simulation that unveils some interesting features of the Polymer BKL
map~\eqref{BKL-map-properties:BKL-map}. The main points of this
simulation are very simple:
\begin{itemize}
\item An initial couple of values for $(u,Q)$ is chosen inside the
  dominion $u\in[1,\infty)$ and $Q\in[0,0.96]$.
\item We remember from Sec.\ref{sec:bianchi-ix}
  and~\ref{sec:parametrization} that, at the end of each Kasner era,
  the $u$ parameter becomes smaller than 1 and needs to be remapped to
  values greater than 1. This marks the beginning of a new Kasner
  era. This remapping is performed through the relations listed on
  page~\pageref{parametrization:remapping}. In the standard case,
  because the standard BKL map is just $u \rightarrow u - 1$, the $u$
  parameter cannot, for any reason, become smaller than 0. For the
  Polymer Bianchi IX, however, it is possible for $u$ to become less
  than zero. This is why we have derived many remapping relations, to
  cover the whole real line.
\item Many values ($\approx 2^{18}$) for the initial conditions,
  randomly chosen in the interval of definition, were tested. We
  didn't observe any ``anomalous behavior'' in any element of the
  sample. This meaning that all points converged asymptotically to the
  standard BKL map as is discussed in the following.
\end{itemize}
                
\begin{figure}
  \begin{center}
    \includegraphics[width=\columnwidth]{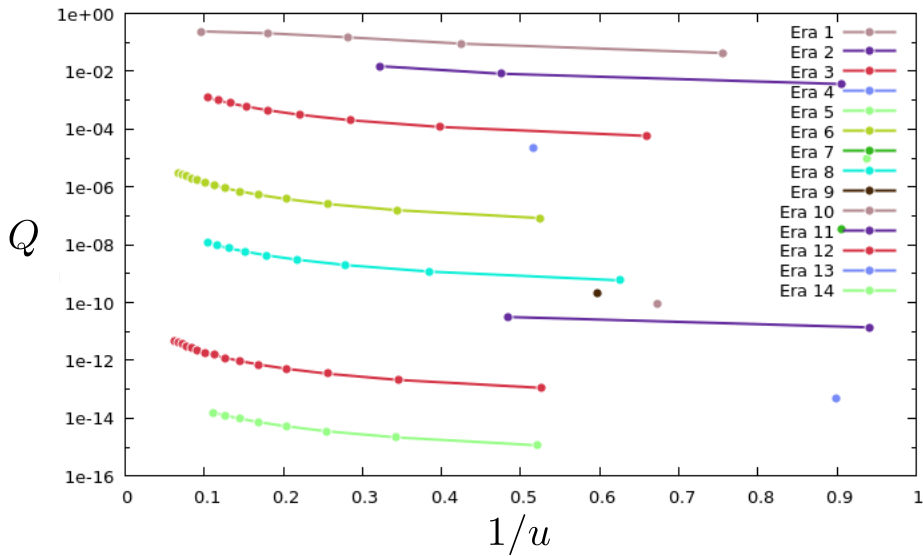}
  \end{center}
  \caption{In this figure the graphical results of a
    sample of the numerical simulation are portrayed. The polymer BKL
    map~\eqref{BKL-map-properties:BKL-map} has been evolved starting
    from the initial value $(u=10,Q=0.24)$. Every point is the result
    of a single iteration of the BKL map. The role played by the
    Polymer BKL map~\eqref{BKL-map-properties:BKL-map} is just to take
    a single point and map it to the next one. A total of 14 Kasner
    eras are shown, each with a different color.  The fact that,
    starting from the upmost point, the map on $Q$ is almost always
    decreasing is very evident from the downwards-going ladder-like
    shape of the eras.}
  \label{fig:numerical-simulation:logscale}
\end{figure}

One sample of the simulation is displayed in
Fig.~\ref{fig:numerical-simulation:logscale} using a logarithmic scale
for $Q$ to show how effective is the map in damping high values of
$Q$.
                
From the results of the numerical simulations the following
conclusions can be drawn:
\begin{itemize}
\item The polymer BKL map~\eqref{BKL-map-properties:BKL-map} is ``well
  behaved'' for any tested initial condition $u\in[1,\infty)$ and
  $Q\in[0,0.96]$: the dominion of definition for $u$ and $Q$ are
  preserved.
\item The line $Q = 0$ is an attractor for the Polymer BKL map: for
  any tested initial condition, the Polymer BKL map eventually evolved
  until becoming arbitrarily close to the standard BKL map.
\item The Polymer BKL map on $Q$ is almost everywhere decreasing. It
  can happen that, especially for initial values $Q \approx 1$, for
  very few iterations the map on $Q$ is increasing, but the overall
  behavior is almost always decreasing. The probability to have an
  increasing behavior of $Q$ gets smaller at every iteration. In the
  limit of infinitely many iterations this probability goes to zero.
\item Since the Polymer BKL map~\eqref{BKL-map-properties:BKL-map}
  tends asymptotically to the standard BKL
  map~\eqref{bianchi-ix:BKL-map}, we expect that the notion of chaos
  for the standard BKL map, given for example in~\cite{Barrow:1982},
  can be applied, with little or no modifications, to the Polymer
  case, too (although this has not been proven rigorously).
\end{itemize}

\section{Physical considerations} \label{sec:physical-considerations}

In this section we address two basic questions concerning the physical link 
between Polymer and Loop quantization methods 
and what happens when all the Minisuperspace variables are discrete,
respectively.

First of all, we observe that Polymer 
Quantum Mechanics is an independent approach from Loop Quantum Gravity.
Using the 
Polymer procedure is equivalent to implement 
a sort of discretization of the considered configurational variables. 
Each variable is treated separately, by introducing a suitable graph 
(de facto a one-dimensional lattice structure): 
the group of spatial translations on this graph is a $U(1)$ type, therefore  
the natural group of symmetry underlying 
such quantization method is $U(1)$ too, differently from Loop Quantum Gravity, where the basic group of 
symmetry is $SU(2)$.

It is important to stress that Polymer 
Quantum Mechanics is not unitary connected 
to the standard Quantum Mechanics, 
since the Stone - Von Neumann theorem is broken in 
the discretized representation. 
Even more subtle is that Polymer quantization procedures applied to different
configurational representations of the 
same system are, in general, not unitary related. 
This is clear in the zero order WKB limit of 
the Polymer quantum dynamics, where
the formulations in two sets of variables, 
which are canonically related in the 
standard Hamiltonian representation, are 
no longer canonically connected in the 
Polymer scenario, mainly due to the 
non-trivial implications of the prescription 
for the momentum operator.

When the Loop Quantum Gravity \cite{Cianfrani:2014, Thiemann:2007loop} is applied to
the Primordial Universe, due to the homogeneity 
constraint underlying the Minisuperspace 
structure, it loses the 
morphology of a $SU(2)$ gauge theory
(this point is widely discussed in \cite{Cianfrani:2010ji, Cianfrani:2011wg}) and 
the construction of a kinematical Hilbert space, as well as of the geometrical
operators, is performed by an effective, 
although rigorous, procedure. 
A discrete scale is introduced in the 
Holonomy definition, taken on a square 
of given size and then the curvature 
term, associated to the Ashtekar-Barbero-Immirzi connection, (the so-called 
Euclidean term of the scalar constraint) 
is evaluated on such a square path. 
It seems that just in this step the 
Loop Quantum Cosmology acquires the features of a polymer graph, associated to an 
underlying $U(1)$ symmetry. 
The real correspondence between the two 
approaches emerges in the semi-classical 
dynamics of the Loop procedure \cite{Ashtekar:2002}, which is isomorphic to the zero 
order WKB limit of the Polymer quantum approach. In this sense, the Loop Quantum
Cosmology studies legitimate the implementation of the Polymer formalism to 
the cosmological Minisuperspace. 

However, if on one hand, the Polymer 
quantum cosmology predictions are, 
to some extent, contained in the Loop Cosmology, on the other hand, the former
is more general because it is applicable 
to a generic configurational representation, 
while the latter refers specifically to 
the Ashtekar-Berbero-Immirzi connection variable. 

Thus, the subtle question arises about which 
is the proper set of variables in order that the 
implementation of the Polymer procedure 
mimics the Loop treatment, as well as
which is the physical meaning of different 
Polymer dynamical behaviours in different sets of variables. 
In \cite{Mantero:2018cor} 
it is argued, for the isotropic Universe 
quantization, that the searched correspondence 
holds only if the cubed scale factor is 
adopted as Polymer variable: in fact this choice leads to a critical density
of the Universe which is independent of the scale factor 
and a direct link between the Polymer 
discretization step and the Immirzi parameter is found. 
This result assumes that 
the Polymer parameter is maintained independent of the scale factor,
otherwise 
the correspondence 
above seems always possible. In this 
respect, different choices of the configuration variables when Polymer
quantizing a cosmological system could be 
mapped into each other by suitable 
re-definition of the discretization step 
as a function of the variables themselves. 
Here we apply the Polymer procedure to the 
Misner isotropic variable and not to 
the cubed scale factor, so that different 
issues with respect to Loop Quantum Gravity 
can naturally emerge. 
The merit of the present choice is that 
we discretize the volume of the Universe, 
without preventing its vanishing behavior. 
This can be regarded as an effective procedure 
to include the zero-volume eigenvalue in the system dynamics, 
like it can happen in Loop Quantum Gravity, 
but it is no longer evident in its 
cosmological implementation.

Thus, no real contradiction exists between the 
present study and the Big-Bounce prediction 
of the Loop formulation, since they are 
expectedly two independent physical 
representations of  the same quantum system. As discussed on the
semi-classical level 
in \cite{Antonini:2018gdd}, when using the 
cubed scale factor as isotropic dynamics, 
the Mixmaster model becomes non-singular 
and chaos free, just as predicted in 
the Loop Quantum Cosmology analysis 
presented in \cite{Bojowald:2004}. 
However, in such a representation, the 
vanishing behavior of the Universe volume 
is somewhat prevented \emph{a priori}, 
by means of the Polymer discretization 
procedure.

Finally, we observe that, while in the 
present study, the Polymer dynamics of 
the isotropic variable $\alpha$ preserves 
(if not even enforces) the Mixmaster chaos, 
in \cite{Montani:2013} the Polymer analysis for the anisotropic variables 
$\beta_{\pm}$ is associated to the 
chaos disappearance. 
This feature is not surprising since the 
two approaches have a different physical meaning: 
the discretization of $\alpha$ has to do 
with geometrical properties of the space-time 
(it can be thought as an embedding variable), 
while the implementation of the Polymer 
method to $\beta_{\pm}$ really affects 
the gravitational degrees of freedom of 
the considered cosmological model.

Nonetheless, it becomes now interesting to understand what happens to the
Mixmaster chaotic features when the two 
sets of variables ($\alpha$ and $\beta_\pm$) are simultaneously Polymer 
quantized. 
In the following subsection, we provide an answer to such an intriguing
question, at least on the base of the semi-classical dynamics.

\subsection{The polymer approach applied to the whole Minisuperspace}

\begin{figure}
 	\begin{center}
                \includegraphics[width=1\columnwidth]{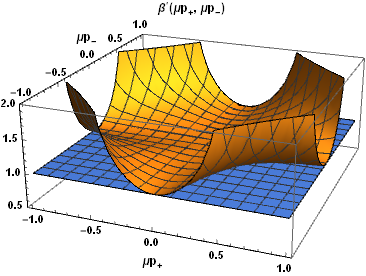}
        \end{center}
        \caption{Anisotropy velocity \eqref{beta'-general-case} (when $\mu_\alpha=\mu$) as a function of $(\mu p_+,\mu p_-)$ 
			 compared with the standard anisotropy velocity $\beta'=1$.}
        \label{fig:anisotropy-velocity}
\end{figure}

According to the polymer prescription \eqref{chiara:polymerSubstitution}, 
the super - Hamiltonian constraint is now: 
\begin{small}
	\begin{multline}
		\mathcal{H}=\frac{B\kappa}{3(8\pi)^2}e^{-3\alpha}
		\left(-\frac{1}{\mu_\alpha^2}\sin^2(\mu_\alpha p_\alpha)+\frac{1}
		{\mu^2}\sin^2(\mu p_+)+\right.\\
		\left.+\frac{1}{\mu^2}\sin^2(\mu p_-)+\frac{3(4\pi)^4}{\kappa^2}
		e^{4\alpha}V_B(\beta_\pm)\right)=0
	\end{multline}
\end{small}
where $\mu_\alpha$ is the polymer parameter associated to $\alpha$ and $\mu$ the one associated 
to $\beta_\pm$. 
The dynamics of the system can be derived through the Hamilton's equations
\eqref{chiara:HamiltonEqs}.
As usual, we start considering the simple Bianchi I case $V_B=0$.
Following the procedure of Sec. \ref{sec:polymer-bianchi-i-dynamics}, we find that the Universe 
is still singular. In fact, the solution to Hamilton's equations in the \eqref{polymer-bianchi-i:time-gauge} time gauge
(with $\mu_\alpha$ instead of $\mu$) is:
\begin{subequations}
	\begin{align}	
		\alpha(t) &=\frac{1}{3}\ln(t) \label{alpha-vs-t}\\
		\beta_\pm(t) &=-\frac{1}{3}\frac{\sin(2\mu p_\pm)}{\sin(2\mu_\alpha p_\alpha)}\ln(t)
	\end{align}
\end{subequations}
from which it follows that $\alpha(t)\rightarrow-\infty$ as $t\to0$.
Moreover, the sum of the squared Kasner indices, calculated by making a reverse canonical transformation
from Misner to ordinary variables, reads as:
\begin{small}
	\begin{equation}
		p_l^2+p_m^2+p_n^2=\frac{1}{3}\left(1+2\frac{\sin^2(2\mu p_+)+\sin^2(2\mu p_-)}{\sin^2(2\mu_\alpha p_\alpha)}\right)
	\end{equation}
\end{small}
from which is possible to derive the anisotropy velocity, according to Eq. \eqref{kasner-vs-anisotropy-velocity}).
The anisotropy velocity can be also calculated through the ADM reduction 
method, as in Sec. \ref{sec:semiclassical-analysis}. 
The new reduced Hamiltonian is:
\begin{equation}
	H_{poly}=\frac{1}{\mu_\alpha}\arcsin\left(\sqrt{\frac{\mu_\alpha^2}{\mu^2}
	(\sin^2(\mu p_+)+\sin^2(\mu p_-))}\right)
\end{equation}
and from the Hamilton's equation for $\beta_\pm$: 
\begin{small}
	\begin{equation}
		\beta'_\pm\equiv\frac{d\beta_\pm}{d\alpha}=\frac{\sin(\mu p_\pm)		
		\cos(\mu p_\pm)}{\sqrt{\left[1-\frac{\mu_\alpha^2}{\mu^2}
		(\sin^2(\mu p_+)+\sin^2(\mu p_-))\right](\sin^2(\mu p_+)+\sin^2(\mu p_-))}}
	\end{equation}
\end{small}
we derive:
\begin{small}
	\begin{multline}\label{beta'-general-case}
		\beta'\equiv\sqrt{\beta'^2_++\beta'^'_-}=\\
		=\sqrt{\frac{\sin^2(\mu p_+)\cos^2(\mu p_+)+\sin^2(\mu p_-)
		\cos^2(\mu p_-)}{\left[1-\frac{\mu_\alpha^2}
		{\mu^2}(\sin^2(\mu p_+)+\sin^2(\mu p_-))\right](\sin^2(\mu p_+)+
		\sin^2(\mu p_-))}}
	\end{multline}
\end{small}
that turns out to be greater than one when $\mu_\alpha\geq\mu$ (see Fig. \ref{fig:anisotropy-velocity}).
Finally, it should be noted that since in the general $V_B\neq0$ case the wall
velocity $\beta'_{wall}$ is not influenced by the introduction 
of the polymer quantization, from Eq. \eqref{beta'-general-case} we find that
the maximum incident angle for having a bounce
against a given potential wall $\left(\theta_{max}=\arccos\left(\frac{\beta'_{wall}}{\beta'}\right)\right)$
is, also in this case, always greater than $\pi/3$ when $\mu_\alpha\geq\mu$.
We can then conclude that also when  
the polymer approach is applied to all the three configuration variables 
simultaneously, the Universe can be singular and chaotic just such as the one analyzed above.

\section{Conclusions}\label{sec:conclusions}

In the present study, we analyzed the Mixmaster model in the framework
of the semiclassical Polymer Quantum Mechanics, implemented on the
isotropic Misner variable, according to the idea that the cut-off
physics mainly concerns the Universe volume.

We developed a semiclassical and quantum dynamics, in order to
properly characterize the structure of the singularity, still present
in the model. The presence of the singularity is essentially due to
the character of the isotropic variable conjugate momentum as a
constant of motion.

On the semiclassical level we studied the system evolution both in the
Hamiltonian and field equations representation, generalizing the two
original analyses in \cite{Misner:1969} and \cite{Belinskii:1970},
respectively.  The two approaches are converging and complementary,
describing the initial singularity of the Mixmaster model as reached
by a chaotic dynamics, that is, in principle, more complex than the
General Relativity one, but actually coincides with it in the
asymptotic limit.

This issue is a notable feature, since in Loop Quantum Cosmology is
expected that the Bianchi IX model is chaos-free \cite{Bojowald:2004,
        Bojowald:2004ra} and it is well known \cite{Ashtekar:2001xp} that
the Polymer semiclassical dynamics closely resembles the average
feature of a Loop treatment in the Minisuperspace.  However, we stress
that, while the existence of the singularity in Polymer Quantum
Mechanics appears to be a feature depending on the nature of the
adopted configuration variables, nonetheless the properties of the Poincar\'e map of the
model is expected to be a solid physical issue, independent on the
particular representation adopted for the system.

The canonical quantization of the Mixmaster Universe that we performed
in the full Polymer quantum approach, i.e. writing down a
Wheeler-DeWitt equation in the momentum representation, accordingly to
the so-called Continuum Limit discussed in \cite{Corichi:2007pr}, is
completely consistent with the semiclassical results.  In fact, the
Misner demonstration, for the standard canonical approach, that states
with high occupation numbers can survive to the initial singularity,
remains still valid in the Polymer formulation, here presented. This
issue confirms that the cut-off we introduced in the configuration
space on the isotropic Misner variable does not affect the
cosmological character of the Mixmaster model.

This result appears rather different from the analysis in
\cite{Montani:2013}, where the polymer approach has been addressed for
the anisotropic Misner variables (the real degrees of
freedom of the cosmological gravitational field) with the emergence of
a non-chaotic cosmology.  Such a discrepancy suggests that the Polymer
regularization of the asymptotic evolution to the singularity produces
more profound modifications when it touches physical properties than 
geometrical features. Actually, the isotropic Misner variable can
be suitably interpreted as a good time variable for the system (an
embedding variable in the language of
\cite{Isham:1985,Isham:1985bis}), while the Universe anisotropies
provide a precise physical information on the considered cosmological
model.

Despite this consideration about the gauge-like nature of the Misner
isotropic variable, which shed light on the physics of our
dynamical results, nonetheless we regard as important to perform
further investigations on the nature of the singularity when other
variables are considered to characterize the Universe volume, since we
expect that, for some specific choice the regularization of the
Big-Bang to the Big-Bounce must take place (see for instance \cite{Antonini:2018gdd}).  However, even on the
basis of the present analysis, we suggest that the features of the Poincar\'e map of
the Bianchi IX model and then of the generic cosmological singularity
(locally mimicked by the same Bianchi IX-like time evolution) is a
very general and robust property of the primordial Universe, not
necessarily connected with the existence of a cut-off physics on the
singularity.

\begin{acknowledgements}
G.P.\ would like to thank Evgeny Grines\footnote{Lobachevsky State
        University of Nizhny Novgorod, Department of Control Theory and
        Dynamics of Systems} for the valuable assistance in determining some of
the mathematical properties of the Polymer BKL map.

We would also like to thank Eleonora Giovannetti for her contribution to the analysis of Sec. \ref{sec:physical-considerations}
\end{acknowledgements}

\bibliography{article.bib}

\end{document}